%% file: main_ICDE.tex
\PassOptionsToPackage{dvipsnames}{xcolor} 

\documentclass[conference]{IEEEtran}
\IEEEoverridecommandlockouts
\usepackage{cite}
\usepackage{amsmath,amssymb,amsfonts,amsthm}
\usepackage{textcomp}
\usepackage{xcolor}
\def\BibTeX{{\rm B\kern-.05em{\sc i\kern-.025em b}\kern-.08em
    T\kern-.1667em\lower.7ex\hbox{E}\kern-.125emX}}
    
 \usepackage{multicol}
\usepackage{adjustbox}

\usepackage{algorithm}
\usepackage{algorithmicx}
\usepackage{algpseudocode}

\usepackage{graphicx}    
\usepackage{caption}     
\usepackage{subcaption}  

\usepackage{booktabs} 
\usepackage{hyperref}

\DeclareMathOperator*{\argmax}{argmax}

\newcommand{\patrick}[1]{{#1}}

\newcommand{\vero}[1]{\textcolor{black}{#1}}

\newtheorem{example}{Example}
\newtheorem{definition}{Definition}

\begin{document}

\title{
Extracting node comparison insights for the interactive exploration of property graphs
}

\author{\IEEEauthorblockN{Cristina Aguiar}
\IEEEauthorblockA{\textit{ICMC} \\
\textit{University of S\~ao Paulo}\\
S\~ao Carlos, Brazil \\
0000-0002-7618-1405}
\and
\IEEEauthorblockN{Jacques Chabin}
\IEEEauthorblockA{\textit{LIFO~UR4022} \\
\textit{Université d'Orléans, INSA CVL}\\
Orléans, France \\
0000-0003-1460-9979}
\and
\IEEEauthorblockN{Alexandre Chanson}
\IEEEauthorblockA{\textit{LIFAT} \\
\textit{University of Tours}\\
Tours, France \\
0000-0001-9195-5950}
\and
\IEEEauthorblockN{Mirian Halfeld-Ferrari}
\IEEEauthorblockA{\textit{LIFO~UR4022} \\
\textit{Université d'Orléans, INSA CVL}\\
Orléans, France \\
0000-0003-2601-3224}
\and
\IEEEauthorblockN{Nicolas Hiot}
\IEEEauthorblockA{\textit{LIFO~UR4022} \\
\textit{Université d'Orléans, INSA CVL}\\
Orléans, France \\
0000-0003-4318-4906}
\and
\IEEEauthorblockN{Nicolas Labroche}
\IEEEauthorblockA{\textit{LIFAT} \\
\textit{University of Tours}\\
Tours, France \\
0000-0002-2794-2124}
\and
\IEEEauthorblockN{Patrick Marcel}
\IEEEauthorblockA{\textit{LIFO~UR4022} \\
\textit{Université d'Orléans, INSA CVL}\\
Orléans, France \\
0000-0003-3171-1174}
\and
\IEEEauthorblockN{Ver\'onika Peralta}
\IEEEauthorblockA{\textit{LIFAT} \\
\textit{University of Tours}\\
Tours, France \\
0000-0002-9236-9088}
\and
\IEEEauthorblockN{Felipe Vasconcelos}
\IEEEauthorblockA{\textit{ICMC} \\
\textit{University of S\~ao Paulo}\\
S\~ao Carlos, Brazil \\
0009-0007-7922-9746}
}


\maketitle

\begin{abstract}
    While scoring nodes in graphs to understand their importance (e.g., in terms of centrality) has been investigated for decades, comparing nodes in  property graphs based on their properties has not, to our knowledge, yet been addressed. In this paper, we propose an approach to automatically extract comparison of nodes in property graphs, to support the interactive exploratory analysis of said graphs. We first present a way of devising comparison indicators using the context of nodes to be compared. Then, we formally define the problem of using these indicators to group the nodes so that the comparisons extracted are both significant and not straightforward. We propose various heuristics for solving this problem. Our tests on real property graph databases show that simple heuristics can be used to obtain insights within minutes while slower heuristics are needed to obtain insights of higher quality.
\end{abstract}


\input{intro}

\input{background}
\input{indicators}
\input{problem}
\input{algorithm}
\input{tests}

\input{related}

\input{conclusion}
\input{ai}

\bibliographystyle{abbrv}
\bibliography{biblio,tkde}

\end{document}

%% file: intro.tex
\section{Introduction}

Imagine an analyst exploring a property graph interactively in order to uncover meaningful patterns and explanations. 
In this work, we address the issue of automatically generating insights into node comparisons in property graphs.
We define a node comparison insight as a pair of contextually similar nodes for which a set of numerical indicators, derived from node and edge properties or the graph topology, demonstrate significant   differences.
These indicators help analysts focus on the most informative contrasts, thus supporting a more guided and efficient exploration process.
To the best of our knowledge, this is the first work to formally define and address the problem of generating comparison insights in property graphs.



 \begin{figure}
     \centering
     \includegraphics[width=8cm]{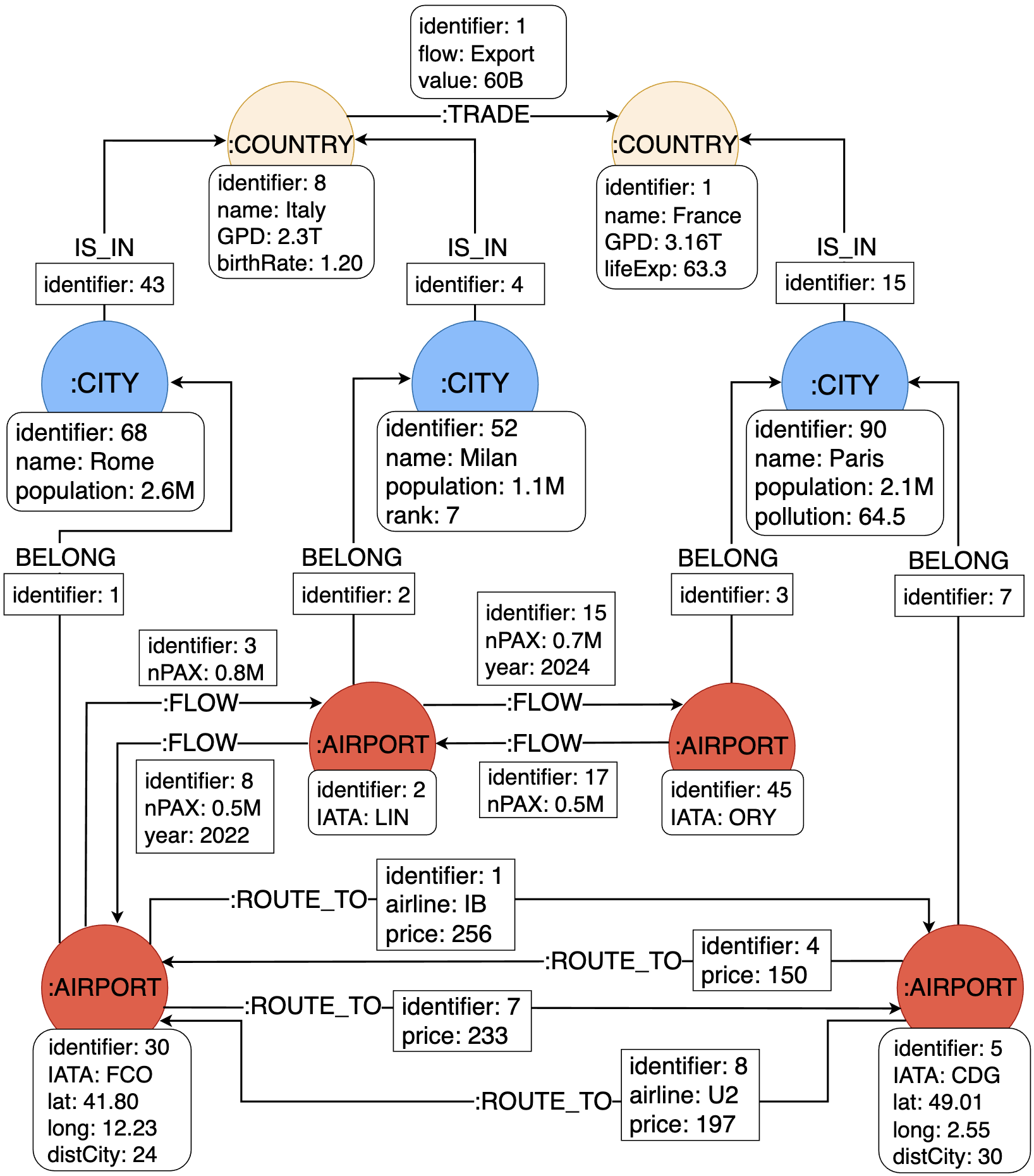}
     \caption{Property graph $G$ of Example \ref{ex:running}. }      \label{fig:ex1-cristina}
 \end{figure}

 \begin{example}
\label{ex:running}
Consider the property graph of Figure~\ref{fig:ex1-cristina} about air routes and trades among countries.
This figure illustrates an extract from our test database, where nodes represent entities and edges correspond to the relationships between them.
Light orange nodes, positioned at the top, denote countries, characterized by numerical properties such as \textit{GDP} and \textit{birthRate}. Blue nodes, located in the middle, represent cities, with numerical properties including the population (in millions) and the pollution index. Red nodes, displayed at the bottom, correspond to airports, described by numerical properties such as latitude and longitude, together with identifiers such as the IATA code.

Edges capture different kinds of relationships. The relationship \texttt{FLOW} indicates passenger exchanges between two airport nodes and includes the property \textit{nPAX}, which specifies the number of passengers. The relationship \texttt{ROUTE\_TO} records all forms of transport, whether goods or people, between airports.

In Figure~\ref{fig:ex1-cristina}, the example comprises two countries (Italy and France), three cities (Roma, Milano, and Paris), and four airports (FCO, LIN, CDG, and ORY).

Assume we are interested in extracting significant comparisons between airports from this graph. To this end, it is necessary to define a set of indicators that capture comparable and meaningful aspects on which airports can be evaluated. Such indicators may rely both on structural properties of the graph and on attributes of the connected entities. For example, one  ($i_1$) can consider the number of distinct routes linking an airport to neighboring airports, obtained by counting the different IATA codes connected to it. Another relevant indicator ($i_2$) is the average price of a route to a neighboring airport, since the relationship \texttt{ROUTE\_TO} includes a property specifying the price of the connection. Additional indicators can be derived from  information concerning the position of the node in the graph, such as the population of the city in which the airport is located  ($i_3$), or the GDP of the country to which it belongs  ($i_4$). These indicators provide a basis for  multi-faceted comparisons among airports.

Table~\ref{tab:indicators1} reports an example of the results obtained for the four indicators mentioned above, computed over an entire database rather than restricted to the extract illustrated in Figure~\ref{fig:ex1-cristina}.\qed

\end{example}


\begin{table}[h]
\centering
\begin{tabular}{lrrrr}
\toprule
\textbf{Airport} & \multicolumn{4}{c}{\textbf{Indicators}} \\
\cmidrule(lr){2-5}
 & $i_1$ & $i_2$€ & $i_3$(M) & $i_4$(T) \\
\midrule
\texttt{CDG} & 285 & 525 & 2.1 & 3.16 \\
\texttt{FCO} & 242 & 450 & 2.6 & 2.30 \\
\texttt{LIN} &  51 & 425 & 1.1 & 2.30 \\
\texttt{ORY} & 178 & 400 & 2.1 & 3.16 \\
\bottomrule
\end{tabular}
\caption{Examples of indicators for the dataset of Example~\ref{ex:running}}
\label{tab:indicators1}
\end{table}

Once indicators are computed, 
to find interesting comparisons, an analyst would probably first normalize the data, keep only one indicator among those  that are correlated, group airports that are similar, and then compare airports within each group. In the example, FCO, CDG, and ORY could be grouped together since they are close in terms of population, leaving LIN in another group.
Among FCO, CDG and ORY, the analyst would report that, while the airports
are in the same group, there are differences in terms of routes 
and
passengers.

Finding these comparisons automatically raise the following questions:
\begin{itemize}
    \item What indicators of comparison should be used?
    \item How to extract them from the graph?
    \item How to split indicators between those used for grouping and those used for comparing?
\end{itemize}




We propose to answer these questions with the following contributions:
\begin{itemize}
    \item a way of devising indicators based on the context of nodes in the graph. This context is formally defined through the relationships between node types and the cardinalities of these relationships, as detected on the graph instance,
    \item a first categorization of indicators to compare nodes in property graphs,
    \item the formal definition of the problem of choosing indicators to  compare nodes and indicators to group nodes. While this problem resembles clustering, its objective is different in the sense that clusters should maximize the significance of intra-cluster comparisons,
    \item various heuristic approaches to solve the aforementioned problem,
    \item series of tests over real graphs to assess the efficiency and effectiveness of the proposed heuristics, 
    \item a benchmark data set based on open data, an excerpt of which is used in Example \ref{ex:running} above, to represent a variety of properties (in nodes and edges) used for devising comparison indicators.
\end{itemize}

The outline of the paper is as follows.
Section \ref{sec:background} presents background definitions.
Section \ref{sec:indicators} details our approach to devise comparison indicators, and
Section \ref{sec:problem} formalizes the problem of extracting comparison insights using these indicators.
Section \ref{sec:algo} introduces our algorithmic solutions to solve these problems.
Section \ref{sec:tests} presents the tests done.
Section \ref{sec:related} discusses related work, while
Section \ref{sec:conclusion} concludes and draws perspectives.

%% file: background.tex
\section{Background}\label{sec:background}

\newcommand{\type}{\textsf{ElemType}}
\newcommand{\Inst}{\textsf{Inst}}


We build on the definitions in
\cite{DBLP:conf/edbt/AnglesBGV25,DBLP:journals/pacmmod/AnglesBD0GHLLMM23} to define property graphs and their schemas, and to introduce the notion of relationship cardinality
with respect to its endpoints.

\begin{definition}[Property graph]
\label{def-PG}
Let $\mathcal{O}$  be a set of identifiers, $\mathcal{L}$ be a set of labels, $\mathcal{P}$ be 
a set of property names and $\mathcal{V}$ be a set of values.
A property graph is defined as a tuple $G = (N,E,\rho,\lambda,\nu)$ where:
\begin{itemize}
    \item $N\subset \mathcal{O}$ is a finite set of node identifiers;
    \item $E\subset \mathcal{O}$ is a finite set of edge identifiers, $N\cap E=\emptyset$
    \item $\rho : E \rightarrow N\times N$ is a total function that maps edges to ordered pairs of nodes (the endpoints of the edge). We denote the source and target nodes of an edge in $E$ by  the functions $\rho^{src}$ and $\rho^{tgt}$, respectively ;
    \item $\lambda : N\cup E \rightarrow \mathcal{L}$ is a total function that assigns labels to nodes and edges;
    \item $\nu :  (N \cup E ) \times \mathcal{P} \rightharpoonup  \mathcal{V}$ is a partial function that defines properties for nodes and edges.
\end{itemize}

An edge in $G$ defines a \textit{relationship} between the two nodes it links. 

To simplify notations we assume that, for all property graphs, for edges having the same label, the source and target nodes also coincide in labels. Formally, the following property holds: 
$\forall e \in E, \forall  e'\in E $ if  $\lambda(e) = \lambda(e')$ then
$\lambda(\rho^{src}(e)) = \lambda(\rho^{src}(e'))$ and  $\lambda(\rho^{tgt}(e)) = \lambda(\rho^{tgt}(e'))$.
\qed

\end{definition}

According to Definition~\ref{def-PG}, in the property graph used as our running example, edges connecting an airport to a city and those connecting a city to a country must have distinct labels.

\begin{example}
Let $G=(N,E,\rho,\lambda,\nu)$ be the graph of Example \ref{ex:running}.
Consider two nodes $n_1$ and $n_2$ describing respectively the airport CDG and the city of Paris, and an edge $e_1$ stating that CDG is an airport that belongs to Paris.
It is: $n_1,n_2 \in N$, $e_1 \in E$,
$\rho(e_1)=(n_1,n_2)$, $\lambda(n_1)=Airport$, 
$\lambda(n_2)=City$, $\lambda(e_1)=BELONG$, 
$\nu(n_1,IATA)=CDG$, 
$\nu(n_2,name)=Paris$, 
$\nu(n_2,population)= 2.1M$.
\end{example}

We also need to consider the schema (or type) of property graphs.
To this end, we begin by introducing the notion of \textit{formal base types}.
A formal base type $T$ is a pair $(l,P)$ where $l \in \mathcal{L}$ and $P \subseteq \mathcal{P}$.
We write $\mathcal{T}$ for the set of all formal base types.



\begin{definition}[Property graph type]
\label{def-PGtype}
A property graph type (graph type in what follows) is a tuple
$S=(N_S, E_S, \type_S, \rho_S)$ where:

\begin{itemize}
    \item $N_S$ and $E_S$ are disjoint finite sets of node and edge type names;
    \item $\type_S: N_S \cup E_S \to \mathcal{T}$ maps elements (node or edge) type names to one formal base type;
    \item $\rho_S : E_S \rightarrow N_S\times N_S$ is a total function that maps edges to ordered pairs of nodes (the endpoints of the edge). We denote the source and target nodes (in $N_S$) of an edge (in $E_S$) by the functions $\rho_S^{src}$ and $\rho_S^{tgt}$, respectively.\qed
\end{itemize}
\end{definition}


\begin{example}
\label{ex:typeG}



\begin{figure}[t]
    \centering
    
    \begin{subfigure}{\textwidth}
       \hspace{0.5cm} \includegraphics[width=7cm]{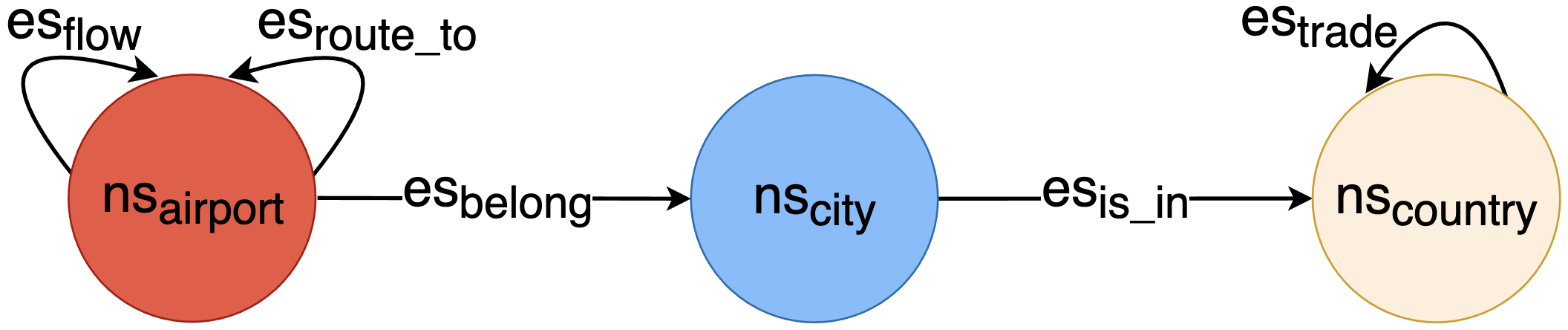}
    \end{subfigure}
    
    \vspace{0.5em} 
    
    \begin{subfigure}{\textwidth}
        \footnotesize
        \begin{tabular}{lll}
        type name    &  label &  set of attributes \\
        \hline 
        $ns_{airport}$ & :AIRPORT & identifier, IATA, lat, long, distCity \\
        $ns_{city}$ & :CITY & identifier, name, population, pollution, rank \\
        $ns_{country}$ & :COUNTRY & identifier, name, GPD, birthRate, lifeExp \\
        $es_{belong}$ & :BELONG & identifier \\
        $es_{is_{in}}$ & :IS\_IN & identifier \\
        $es_{flow}$ & :FLOW & identifier, nPAX, year \\
        $es_{route_{to}}$ & :ROUTE\_TO & identifier, airline, price \\
      $es_{trade}$ & :TRADE & identifier, flow, value \\    
      \end{tabular}
    \end{subfigure}
    \caption{Property graph type $S$ for Example~\ref{ex:running}
    \label{fig:img-table}}
\end{figure}


In  Figure~\ref{fig:img-table},  we show a  graph type $S_1$ in the context of  our running example.
It defines three node types, i.e., $N_S = \{ns_{airport}, $ $ns_{city}, $ $ns_{country}\}$,   and five edge types, i.e., 
$E_S=\{es_{trade}$, $es_{is\_in}$, $es_{belong}$, $es_{flow}$, $es_{route\_to}\}$.
Let us consider $es_{belong}$ and its associated nodes. We have:
$\type_S(ns_{airport})$=(:AIRPORT, \{identifier, IATA, lat, long, distCity\}  ),
$\type_S(ns_{city})$=(:CITY, \{identifier, name, population, pollution, rank\}),
$\type_S(es_{belong})$ =(:BELONG, \{identifier\}),
$\rho_S(es_{belong})=(ns_{airport},ns_{city})$.\qed
\end{example}

\begin{definition}[Element validity and type instance]
Given a property graph $G$, an element $x$ (either a node or an edge) in $G$, with label $l_1$ and a set of property names $P_1$,
is said to be \emph{valid} with respect to a formal base type $T = (l, P)$ if $l_1 = l$ and $P_1 \subseteq P$.
We also say that such an element is an instance of $T$ (denoted by $x \in Inst(T)$).\qed
\end{definition}



\begin{definition}[Instance of a graph type]
Given a  graph type $S = (N_S, E_S, \type_S,\rho_S, )$ and a property graph $G = (N, E, \rho, \lambda, \nu)$, 
we say that $G$ is an instance of $S$ (denoted by $G \in \Inst(S)$) if: 
\begin{itemize}
    \item for every node $n \in N$ there exist 
    $n_S \in N_S$ such that $n$ is valid with respect to $\type_S(n_S)$ and 
    \item for every edge $e \in E$:
    \begin{itemize}
        \item there exist 
        $e_S \in E_S$ such that $e$ is valid with respect to $\type_S(e_S)$;
        \item $\rho^{src}(e)$ is valid with respect to $\type_S(\rho_S^{src}(e_S))$ and 
        \item $\rho^{tgt}(e)$ is valid with respect to $\type_S(\rho_S^{tgt}(e_S))$.
    \end{itemize}
\end{itemize} 
We also say that $G$ is valid with respect to $S$.\qed
\end{definition}

\begin{example} The graph in Figure~\ref{fig:ex1-cristina} is an instance of  the  graph  type of Figure~\ref{fig:img-table}. Nodes such as the ones representing the airports CDG and LIN have different property sets, yet both satisfy the properties defined for the node type $ns_{airp}$, ensuring their validity in the instance.\qed
\end{example}

We next define the \textit{cardinality of a relationship with respect to  its endpoints}, by examining all its instances in a given graph instance.
Consider the graph type $S$ in Figure~\ref{fig:img-table} and its valid instance $G$ in Figure~\ref{fig:ex1-cristina}.
To determine the cardinality of $es_{belong}$ with respect to its source node $ns_{airport}$ in $S$, we count, for each instance $n$ of $ns_{airport}$ in $G$, the number of outgoing $BELONG$ relationship instances from $n$.
In $G$,   each airport belongs to exactly one city, so $\mathit{card}^{src} = 1$.
Similarly, to determine the cardinality of $es_{belong}$ with respect to its target node $ns_{city}$ in $S$, we count, for each instance $n$ of $ns_{city}$ in $G$, the number of ingoing $BELONG$ relationship instances to  $n$.
In $G$,    since Paris has two airports, we have $\mathit{card}^{tgt} = *$.
The following definition formalises this notion.


\begin{definition}[Relationship cardinalities with respect to its endpoints]
Let $G = (N, E, \rho, \lambda, \nu)$ be a property graph valid with respect to a graph type $S= (N_S, E_S, $ $\type_S,$ $\rho_S)$.
Given $e_S \in E_S$,

\begin{itemize}
    \item if $\exists n \in \Inst(\type_S(\rho_S^{src}(e_S))$)
    such that
    
    $|\{e \in \Inst(\type_S(e_S))\mid \rho^{src}(e) = n\}| > 1$
    then 
    
    $card^{src}(e_S) = *$ else $card^{src}(e_S) = 1$
    \item if $\exists n \in \Inst(\type_S(\rho_S^{tgt}(e_S))$) such that 
    
    $|\{e \in \Inst(\type_S(e_S)) \mid \rho^{tgt}(e) = n\}| > 1$ then
    
    $card^{tgt}(e_S) = *$ else $card^{tgt}(e_S) = 1$. \qed
\end{itemize}

\end{definition}

We now introduce our notation for paths on a graph.

\begin{definition}[Path]
Given a graph $G = (N, E, \rho, \lambda, v)$, we define a path as a sequence of nodes and edges of the form 
$p = (n_1, \alpha_1, ..., \alpha_k, $ $n_{k+1})$ where
\begin{itemize}
\item for $k=0$ (empty path), $p=(n_1)$, $\lambda(p) = \epsilon$;
\item for $k > 0, $
\begin{itemize}
\item $\forall i \in [1, k], n_i \in N$, $\alpha_i \in \{ e_i, e_i^{-1}\}$, $e_i \in E$ and $\rho(e_i) = (n_i, n_{i+1})$;
\item $e^{-1}$ is used  to indicate that an edge $e$ is traversed in its reverse direction  and 
\item $\lambda(p)=(\lambda(\alpha_1).\ldots.\lambda(\alpha_k))$.
\end{itemize}
\end{itemize}
For simplicity, a path $p$ between nodes $n,n'$ will be noted $n \xrightarrow{\lambda(p)} n'$.
Paths over graph types are defined analogously.

When $G$ is valid with respect to a graph type $S$, each path $n \xrightarrow{\lambda(p)} n'$ in $G$ corresponds to a path $p_S= n_S \xrightarrow{\lambda(p)} n'_S$ in $S$:
 for each node $n_i$ and each edge $e_i$ (or $e^{-1}_i$)  in $p$ there exists $e_{i_{S}} \in E_S$ such that  $e_i \in \Inst(e_{i_S})$, 
 $n_i \in \Inst(\rho^{src}(e_{i_{S}}))$ and   $n_{i+1} \in \Inst(\rho^{tgt}(e_{i_{S}}))$ (or the reverse, if the edge is traversed backward).
 Moreover we note $n_S \xrightarrow[1]{\lambda(p)} n'_S$  when 
 $\forall e_{i_{S} }\in p_S$ we have 
 $card^{src}_S (e_{i_S})=1 \vee  card^{tgt}_S  (e^{-1}_{i_S}) =1$;
 otherwise we note 
 $n_S \xrightarrow[*]{\lambda(p)} n'_S$. \qed
\end{definition}


 
%
%
%
%
%
%
%
 
  \begin{example}
  In Figure~\ref{fig:ex1-cristina}, let us call node $n_1$ the one representing Milan and node $n_2$ the one representing Rome, while nodes $n_3$ and $n_4$ correspond to their respective airports.
  Path $p_1= (n_1, $ $BELONG ^{-1}, n_3, FLOW, n_4,BELONG , n_2)$ is a path in $G$, and 
  paths $ns_{City} \xrightarrow[*]{es_{BELONG}^{-1}. es_{FLOW}. es_{BELONG}} ns_{City}$, \linebreak $ns_{Airport} \xrightarrow[1]{es_{BELONG} .es_{IS\_IN}} ns_{Country}$ are paths in $S$.  \qed
  \end{example}
 

%% file: indicators.tex
\section{Devising comparison indicators}\label{sec:indicators}


To the best of our knowledge, the problem of devising  indicators to compare nodes in a property graph has never been addressed.
We adopt an idea from the identification of comparison features in knowledge graphs \cite{DBLP:conf/esws/GiacomettiMS21}.
The idea is first to identify a comparison context for each node to be compared.
Then, candidate indicators are devised based on the properties found in the context, and on the context topology. 

\subsection{Indicator provenance and computation}


Indicators are important because they enable complex graph nodes to be expressed as comparable and meaningful descriptors that support data analysis. By encoding both semantic content and structural characteristics, they ensure that comparisons take into account both the data and the context while remaining transparent and reproducible. However, a key challenge lies in defining the right indicators — deciding which aspects of the data and topology to capture and how to measure them.
Table \ref{tab:indicators} summarizes  different ways of  computing indicators, where we distinguish: 
\begin{itemize}
    \item the provenance: the indicator value comes from a node property, from the graph topology (e.g., in-degree), or from both (e.g., weighted degree, the sum of edge weights when connections carry weights),
    \item scope: the indicator value is from the node itself (inner), intrinsic to the context (i.e., the neighbors) of the node, or global (i.e., relative to the complete graph),
    \item computation: primitive if the indicator value comes directly from a property value  or derived if the indicator value involves a computation (e.g., an aggregate of several values).
\end{itemize}

For instance, the population of an airport’s city is a property-based, contextual, primitive indicator, since its value is the primitive property of population taken from the airport’s context. By contrast, an indicator such as the percentile of shortest paths to the node with the highest PageRank score would be a topology-based, global, derived indicator.

In this paper, we focus on indicators derived from the numerical attributes of a node and its neighbourhood. Here, the neighbourhood is defined by the node’s context. This context encompasses the hierarchy to which the node belongs and its nearest neighbours. In Example~\ref{ex:running}, comparing two airports involves considering the following: (i) the properties of the airport nodes themselves, (ii) the properties of the nodes in the airport's context (i.e. the city and country to which each airport belongs), (iii) the properties of neighbouring airports, and (iv) the properties of the relationships connecting these nodes.
The following section formalizes the notion of context.


\begin{table}[]
    \centering
    \begin{tabular}{llll}
    description & provenance & scope & computation \\
         \hline 
         \hline 
    property of $n$ & property & inner & primitive\\
    \hline
    property of $l$ or $r\in p$ \\
    \hspace*{1cm} with $n \xrightarrow[1 ]{\lambda(p)} l$  & property& context & primitive\\
    \cline{2-4}
     \hspace*{1cm} with $n \xrightarrow[*]{\lambda(p)} l$  & property & context & derived \\
    \hline 
    degree of $n$ for relation $r$ &  topology & context & derived\\
    \hline 
    centrality of $n$  & topology & global & derived
    \end{tabular}
    \caption{Example of indicators for node $n$}
    \label{tab:indicators}
\end{table}

\subsection{Context of a node} 
\label{sec:contextNode}

\input{newProposal}


\subsection{Validating candidate indicators}\label{sec:validating}


Let $n_S$ be a node type and  $i=\langle \lambda(p), prop, type, agg\rangle$ an indicator in $I_{n_S}$, with  $value(n_S,i)$ the set of values $\{value(n,i) | n\in \Inst(\type(n_S))\}$.
\patrick{ 
Let $x_{prop}$ be the element type for which $prop$
is defined, i.e., $n_S$ if $p=\epsilon$,
the last edge or node of $p$ according to type, otherwise.}
$i$ is considered valid with respect to a set $I$ of valid indicators and a set $P$ of property names,
if it satisfies the following desirable validation-properties\footnote{We use the term validation-properties to differentiate from graph properties, simply called properties throughout the paper.} (where $\alpha, \beta$ are constants): 
\begin{itemize}
    \item acceptable variance: $ 1 \leq  \alpha \leq  |value(n_S,i)| \leq \beta \leq | \Inst($ $\type(n_S) )|$ 
   \item acceptable density: 
     $|value(x_{prop},prop)| \geq \gamma $ 
    \item non-redundancy: for some function $f$, $\forall i'\in I, value(n_S,i)\not= f(value(n_S,i'))$ 
    \item contextualization: for some function $f$, $\forall v \in value(n_S,i), v =f(i,C_{n_S})$ 
    \item scaling: $value(n_S,i) \in [0,1]$
    \item discarded properties: for a set $P$ of property names, $prop \not\in P$
\end{itemize}

Acceptable variance allows to discard indicators with too few or too many values, often deemed non informative.
Acceptable density allows to discard indicators with too many null values.
Non-redundancy allows to discard correlated indicators, i.e., indicators whose values can be deduced from other indicators.
Contextualization requires the values of the indicator to be a function of the indicator's location in the context.
This can be used to attenuate the value of properties found using long paths from the nodes to be compared.
Scaling requires the indicators to be normalized, in order not to favor one against the others.
Finally, discarded properties can be seen as user-defined rejection of unwanted properties (e.g., surrogate keys or identifiers, zip codes, latitude/longitude, etc.).
Such validation-properties are often used in machine learning. 



%% file: newProposal.tex

The notion of context is formally defined on a graph type \(S\).  
By abuse of language, we sometimes refer to the context of a node \(n\) in \(G\) when we actually mean the context of the corresponding node \(n_S\) in the graph type \(S\).
To determine the \emph{context} of a node $n_S$  in a graph type \(S\), we proceed as follows:  

\begin{enumerate}
  \item Consider all relationships incident to the corresponding node \(n_S\) in \(S\).
  \item Consider the hierarchical environment of \(n_S\) in \(S\). Concretely, this means collecting all paths of the form
  $    p_S = n_S \xrightarrow[1]{\lambda(p)} n'_S,$
  and adding every element that appears on \(p_S\) to the context of \(n\).
\end{enumerate}

Formally, the context of a node is a subgraph \(S_1 \subseteq S\).  
An instance in \(G\) is a subgraph \(G_1 \subseteq G\), that is valid with respect to \(S_1\).

As an example, in Figure~\ref{fig:img-table}, the context of node $ns_{airport}$ corresponds to the whole graph in Figure~\ref{fig:img-table}, but without the edge \(es_{trade}\). Notice that the nodes \(ns_{City}\) and \(ns_{Country}\) represent the hierarchical environment of 
node $ns_{airport}$.

The following definition formalises the notion of  context.





\begin{definition}[Context of a node]\label{def:context}
Let $S=(N_S, E_S, \type_S, \rho_S)$ be a  graph type.
The context of $n_S \in N_S$ is defined as a sub-graph of $S$, denoted by $C_{n_S} = (
N_{n_S}, E_{n_S}, \type_{n_S}, \rho_{n_S})$ 
where:

\begin{itemize}
    \item  $N_{n_S}=\{n_S\} \cup \{n_S'\in N_S | \exists p, n_S \xrightarrow[1]{\lambda(p)} n_S'\}$
    $\cup ~\{n_S'\in N_S | \exists e_S\in E_S, (n_S,n'_S)=\rho_S(e_S)\}$;
    \item $E_{n_S}=\{e_S\in E_S | \exists p, n_S \xrightarrow[1]{\lambda(p)} n'_S, e_S\in p\}
    \cup \{e_S\in E_S | (n_S,n'_S)=\rho_S(e_S)\}$;
    \item $\type_{n_S}={\type_S}_{|_{E_{n_S}\cup N_{n_S}}}$;
    \item $\rho_{n_S}={\rho_S}_{|_{E_{n_S}}}$ \qed
\end{itemize}
\end{definition}



\subsection{Candidate indicators}

Once the context is defined, we consider as candidate indicators all numerical properties associated with the nodes and edges within this context or derived from them (e.g., the count of neighbours).
The values used for these indicators also depend on the cardinality of the relationships involved.
Without loss of generality, we do not consider indicators of global scope (e.g., centrality), as such  indicators can be pre-computed and recorded as node properties, therefore becoming parts of the node context.


For a relationship $e_S$ with $\mathit{card}^{src}(e_S) = 1$ (or $\mathit{card}^{tgt}(e_S) = 1$ when the edge is traversed in reverse), the property values can be used directly—for example, the population of cities or GDP of countries.
In contrast, for relationships where $\mathit{card}^{src}(e_S) = *$ (or $\mathit{card}^{tgt}(e_S) = *$ when traversed in reverse), it is necessary to aggregate the properties of the neighboring elements—for instance, by computing the average number of employees at neighboring airports, or the average number of passengers sent to or received from them. 

Indicators are numerical properties that support meaningful comparisons in a data graph.
To consider the values of indicators effectively, it is often necessary to examine their traces—that is, to understand  where  in the graph they have been obtained.
To capture this information precisely, we define indicator names as quadruplets that describe the origin of the property being considered, namely,
the path followed to retrieve it, the involved property, whether the property comes from a node or edge, and the involved aggregation function. 

\begin{definition}[Operators on properties]\label{def:aggregation}
Let $p = n_S \xrightarrow{\lambda(p)} n_S'$ be a path on a context $C_{n_S}$. 
Let $prop \in \mathcal{P}$ be a property of the  last node or the last edge in $p$.
Let $\mathcal{A}$ be a set of aggregation functions (typically, \textit{sum}, \textit{average}, \textit{minimum}, \textit{maximum} and \textit{count}) and let $OpDict: \mathcal{P} \xrightarrow{} 2^\mathcal{A}$ be a domain dictionary indicating the aggregation functions that are the most appropriated for each property (e.g., sum for population and average for pollution).

\noindent
We define $Op(p,prop)$ as the set of operators associated with property $prop$ in path $p$, specified as follows:

$Op(p,prop) = \begin{cases}
    \{\textsf{id}\}              
        & \text{if } n_S \xrightarrow[1]{\lambda(p)} n_S'  \text{ or } p=(n_S)     \\
    OpDict(prop) 
        & \text{if } n_S \xrightarrow[*]{\lambda(p)} n_S'   
\end{cases}$

\noindent 
where 
\textsf{id} is the identity function.\qed
\end{definition}






\begin{definition}[Candidate indicators for a context]\label{def:indicators}
Let $C_{n_S} = (N_{n_S}, E_{n_S}, \type_{n_S}, \rho_{n_S})$ be a context for node type $n_S$.
Let $p$ be a path on graph $C_{n_S}$.






\vspace{0.2cm}
\noindent
An indicator name is a quadruplet $\langle path, prop, type, op
\rangle$ where
\begin{itemize}
    \item $path$ corresponds to the label $\lambda(p)$ of a path $p$ or $\epsilon$ for empty path;
    \item $prop \in \mathcal{P}$ is a property name;
    \item $type \in \{\mbox{'node'}, \mbox{'edge'}\}$ indicates  if the property comes from the node type or the edge type leading to it;
    \item  $op \in Op(p,prop)$ is an operator  for the property in the path.

\end{itemize}

%
%


\noindent
For a  path  $ p= (n_S, {e_S}_1, \dots, {e_S}_k, n'_S)$, with $(L_n, P_n) = \type_S(n'_S)$, $(L_e, P_e) = \type_S({e_S}_k)$, 
the set $I^p_{n_S}$ is the union: \\
\indent
    $\{\langle \lambda(p), prop, \mbox{'node'}, op\rangle | 
        prop \in P_n$,  
        $op \in Op(p,prop)\}$ 
$\cup$ \\
\indent
    $\{\langle \lambda(p), prop, \mbox{'edge'}, op\rangle |  
        prop \in P_e$, 
        $op \in Op(p,prop)\}$

\noindent
The set of indicators $I_{n_S}$ for a node type $n_S$ is $ \bigcup_p I^p_{n_S} $, for \textbf{all} path $p = n_S \xrightarrow{\lambda(p)} n'_S $ in $C_{n_S}$.    
    \qed

\end{definition}

\begin{example}
\label{ex:ind1}
Let us consider the context of $ns_{airport}$ (Section~\ref{sec:contextNode}) in the graph type of Figure~\ref{fig:img-table}.  
As indicators for $ns_{airport}$, we cite: \\  
$\bullet$ $i_1 = \langle es_{route\_to}, IATA, \mbox{'node'}, count\rangle$,  which indicates the total number of neighbouring airports.\\
$\bullet$ $i_2 = \langle es_{route\_to}, price, \mbox{'edge'}, avg\rangle$, which indicates the average price of a route to a neighbouring airport. \\ 
$\bullet$ $i_3 = \langle es_{belong}, population, \mbox{'node'}, id\rangle$, which indicates the population of the airport's city. \\  
$\bullet$ $i_4 = \langle es_{belong}. es_{is\_in}, GPD, \mbox{'node'}, id\rangle$, which indicates the GPD  of the airport's country. \qed 
\end{example}

%
%


\begin{definition}[Indicator values]\label{def:values}
Let  $G = (N,E,\rho,\lambda,\nu)$ be a valid graph with respect to graph type $S = (N_S, E_S, \type_S, \rho_S)$.
Let $n_S \in N_S$ be a node type and $I_{n_S}$ be a set of indicators.

\noindent
The indicator value of a node $n \in \Inst(\type_S(n_S))$ and
indicator $i = \langle path_i, prop_i, type_i, op_i\rangle \in I_{n_S}$
is defined as follows.
Let $N_i = \{n' \in N | \exists p= n \xrightarrow{path_i} n'\}$, i.e., the nodes reachable with $path_i$,
and $E_i = \{e' \in E | \exists p= (n,... ,e',n') \wedge p= n \xrightarrow{path_i} n'\}$, i.e., the 
last edges of $path_i$.

\noindent
The value of the indicator $i$ is given by:

\noindent
$\text{value}(n, i) = \begin{cases}
    null  &  \text{if } N_i = \emptyset \\
    \mathrm{op_i}(\{\nu(n', prop_i) | n' \in N_i\})  & \text{if } type_i = \mbox{'node'} \\
    \mathrm{op_i}(\{\nu(e', prop_i) | e' \in E_i\}) & \text{otherwise} 
\end{cases}$ 
\qed

\end{definition}

\begin{example}
For the indicators of Example~\ref{ex:ind1}, if we restrict ourselves to the values shown in Figure~\ref{fig:ex1-cristina}, we obtain the  results in Table~\ref{tab:IndComp}. Recall that Figure~\ref{fig:ex1-cristina} represents only a small excerpt from our dataset and that  Table~\ref{tab:indicators1} shows the same indicators for an entire test dataset.\qed
\end{example}
\begin{table}
\begin{center}
\small
\begin{tabular}{l c c c c c}
\toprule
\textbf{Airport} & \textbf{Node Id} & \multicolumn{4}{c}{\textbf{Indicators}} \\
\cmidrule(lr){3-6}
& & $i_1$ & $i_2$ (€) & $i_3$ (M) & $i_4$ (T)  \\
\midrule
CDG & $30$ & $2$ & $173.5$ & $2.1$  & $3.16$ \\
FCO & $28$ & $0$ & $244.5$ & $2.6$  & $2.3$ \\
LIN & $5$  & $0$ & $null$ & $1.1$ & $2.3$ \\
ORY & $45$ & $2$ & $null$ & $2.1$ & $3.16$ \\
\bottomrule
\end{tabular}
\normalsize
\end{center}
\caption{Indicator values computed from values in Figure~\ref{fig:ex1-cristina}.\label{tab:IndComp}}
\end{table}


\begin{definition}[Collection of indicator values.]
Let $S$ be a graph type $S$ and $n_S$ be a node type of $S$.
\patrick{
The values of property $prop$ for node type $n_S$
is $values(n_S,prop)=\{\nu(n, prop) | n\in \Inst(\type_S(n_S))\}$.
}
The values of indicator $i$ for node type $n_S$ is the union of $value(n,i)$ for all $n \in \Inst(\type_S(n_S))$, 
i.e.,
$value(n_S,i)=\bigcup_{\Inst(\type_S(n_S))} value(n,i)$, 
and the values of all indicators in $I_{n_S}$ for node type $n_S$
is the union of $value(n_S,i)$ for all $i\in I_{n_S}$, i.e., $value(n_S)=\bigcup_{I_{n_S}} value(n_S,i)$. \qed
\end{definition}

%% file: problem.tex
\section{Problem definition}\label{sec:problem}

The previous section detailed how, 
for a given node type, a set of indicators 
can be defined, by collecting 
and aggregating properties
in the context of the node. 
The present section addresses the problem of
extracting comparisons of nodes of a given type, based on this set of indicators.

\subsection{Node comparison insights}

\begin{definition}[Comparison / grouping scheme]
Let  $G = (N,E,\rho,$ $\lambda,\nu)$ be a valid graph with respect to graph type $S = (N_S, E_S, $ $\type_S, \rho_S)$.
Let $n_S \in N_S$ be a node type and $I_{n_S}$ be a set of indicators for $n_S$.
A node comparison scheme (resp., a grouping scheme) is a tuple 
$(n_S,I)$\vero{, where}
$I \subseteq I_{n_S}$ is the set of indicators used for the comparison (resp. the grouping).
\end{definition}

Given a comparison scheme $(n_S,I)$, a comparison is a tuple 
$(n,n',I)$ where  $n,n' \in  \Inst(\type(n_S))$. 
Let  $c=(n,n',I)$ be a comparison and $i$ an indicator in $I$. The \textit{significance} of the comparison $c$ for $i$ is given by a mathematical expression $sig_i$. For instance, $sig_i(n,n')= |value(n,i)-value(n',i)|$. The overall significance of a comparison is given by aggregating (e.g., with sum) the significance for all indicators of $I$.

In what follows, our aim is to maximize the significance of comparisons. However, to disregard comparisons that are straightforward, we require that comparisons be made among nodes that are similar, hence the introduction of grouping schemes. Given $g=(n_S,I)$ a grouping scheme, $i$ an indicator in $I$, and  $n,n' \in  \Inst(\type(n_S))$,
the \textit{distance} between $n$ and $n'$ for $i$ is given by a mathematical expression $dist_i$, e.g., $dist_i=(value(n,i)-value(n',i))^2$. The overall distance for $g$ is given by aggregating the distances for all indicators of $I$.

\begin{definition}[Node comparison insight]
Let  $G = (N,E,\rho,\lambda,\nu)$ be a valid graph with respect to graph type $S = (N_S, E_S, \type_S, \rho_S)$. Let $n_S \in N_S$ be a node type
and  $n,n' \in  \Inst(\type(n_S))$. 
Given a comparison scheme $c=(n_S,I)$ and a grouping scheme $g=(n_S,I')$
with $I\cap I'=\emptyset$, 
a node comparison insight for $n$ and $n'$ is the tuple 
$(n,n',\{(value(n,i),value(n',i)) | i \in I\})$.
\end{definition}

In what follows, we formalize the problem of extracting comparison
insights as the problem of \textit{finding schemes for grouping and comparing nodes},
maximizing the significance of comparison and minimizing the distance between \vero{the}
nodes that are compared.

\subsection{Problem formulation}

Given a set $I_{n_S}$ of indicators for a node type $n_S$ and 
the set \vero{$N'=$} $\Inst(\type(n_S))$ of nodes to be compared, 
the problem of extracting comparison insights  consists of finding
a comparison scheme and a grouping scheme.
In other words, we search among all indicators in $I_{n_S}$, 
a subset $P_S$ such that $(n_S,P_S)$ is a comparison scheme 
and a subset $P_D$, disjoint from $P_S$, such that $(n_S,P_D)$ 
is a grouping scheme.
We do not require that $P_S \cup P_D = I_{n_S}$, meaning that what
we are looking for is actually a 3 partition of $I_{n_S}$,
where $P_U= I_{n_S} \setminus (P_S \cup P_D)$ is the set of indicators
that are used neither for comparisons nor groupings.

Intuitively, a general formulation of the problem could be:
\[
    \argmax \sum_{k=1}^{K} \Bigl(
        \sum_{n,n'\in S_k}\sum_{i\in P_S} sig_i(n,n') -
        \sum_{n,n'\in S_k}\sum_{i\in P_D} dist_i(n,n')
    \Bigr)
\]


\noindent
which can be understood at finding $P_S$ and $P_D$
to compute $K$ clusters of nodes \vero{\{$S_1,...,S_K$\}}, 
by minimizing the intra-cluster distance (second term) while
maximizing the significance inside clusters (first term).

In what follows, we consider a version of the problem where the absolute value of difference is used for the significance of the comparisons and the Euclidean distance is used for the groupings. 
We also weigh each term of the problem to balance both the distribution of indicators
among  clusters and  the distribution of indicators among 
$P_{S}$ and $P_{D}$.

We thus look for \( P= \{P_{S},P_{D},P_{U}\}\) a 3-partition of $I_{n_S}$ that define $K$ clusters of nodes,
yielding the following optimization problem:


\begin{equation}\label{eq:obj}
\begin{split}
    \argmax_{\{S_1,\ldots,S_K\},P} \quad & \sum_{k=1}^{K} \frac{1}{|S_k|} \frac{|P_S|}{|P|} \\
    & \sum_{n,n'\in S_k}\sum_{i\in P_S}\bigl|value(n,i)-value(n',i)\bigr| \quad -\\
    & \sum_{k=1}^{K} \frac{1}{|S_k|} \left(1-\tfrac{|P_D|}{|P|}\right) \\
    & \sum_{n,n'\in S_k}\sum_{i\in P_D}\bigl(value(n,i)-value(n',i)\bigr)^2
\end{split}
\end{equation}

\textbf{such that :} 
\vspace{-1em}

\begin{gather}
    | P_{S} | > 0 \label{eq:noempty-clust}\\
    | P_{D} | > 0 \label{eq:noempty-comp}\\
    P = P_{S} \cup P_{D} \cup P_{U} \label{eq:partition-1}\\
    P_{S} \cap P_{D} = \emptyset, P_{U} \cap P_{D} = \emptyset, P_{S} \cap P_{U} = \emptyset \label{eq:partition-2}\\
    \forall k \in [1,K], |S_k|>1 \label{eq:partition-3}
\end{gather}


\noindent
where the \vero{first term of the objective function computes the significance for indicators in $P_S$ and the second term computes the Euclidean distance for indicators in $P_D$. Remark that the} second term of the objective function corresponds to a classical clustering objective as used by the K-Means method, \( S = \{S_1,\dots,S_K\} \) are clusters of vertices \vero{in N'}, 
and K is a constant integer (the number of clusters). 

Note that although equations (\ref{eq:partition-1}) and (\ref{eq:partition-2}) represent the classical integrity constraints of a 3-partition, equations (\ref{eq:noempty-clust}) and (\ref{eq:noempty-comp}) are needed to ensure there exists at least one indicator for the purpose of clustering (respectively for comparison).

Finally, in equation (\ref{eq:obj}), 
the terms $\frac{|P_S|}{P}$ and  $\frac{|P_D|}{P}$ are used to balance indicators among $P_S$ and $P_D$.


\vero{Further constraints can be added to represent user preferences, indicating that a particular indicator must belong to $P_S$, $P_D$ or $P_U$.}



As this problem is clearly non polynomial, in the following section we propose various heuristics to solve it.


%% file: algorithm.tex
\section{Algorithmic solutions}\label{sec:algo}

This section presents the algorithms we propose to collect indicators and extract comparison insights.

\subsection{Collecting indicators}

Algorithm \ref{algo:collect} details how  to obtain  a set of valid indicators from a property graph, given a node type.

\begin{algorithm}
\begin{algorithmic}[1]
\Require a property graph $G = (N,E,\rho,\lambda,\nu)$ 
valid wrt a graph type $S=(N_S,E_S,\type_S,\rho_S)$, 
a node type $n_S\in N_S$ ,  thresholds $\alpha, \beta, \gamma$,
a set $D$  of property names to discard
\Ensure a set $I$ of valid indicators for $n_S$

\State $I = \emptyset$ 


\State compute context $C_{n_S}$ for $n_s$ 
\Comment{Def. \ref{def:context}}

\State compute the set of indicators $I_{n_S}$ for $n_s$
\Comment{Def. \ref{def:indicators}}

\For{$i=\langle p, prop, type, agg\rangle \in I_{n_S}$} 

\If{$|value(x_{prop},prop)| \geq \gamma$} \Comment{acceptable density}



\If{$i$ is not correlated to any $i'$ in $I$} 

\Comment{non redundancy}

\If{$prop \not\in  D$ and $\alpha \leq |value(n_S,i)| \leq \beta$}

\Comment{acceptable variance} 

\Comment{and discarded properties}

\State scale($i,value(n_S,i)$)\Comment{scaling}
\State contextualize($i,p$)\Comment{contextualization}

\State $I = I \cup i$
\EndIf
\EndIf
\EndIf

\EndFor

\State \Return{$I$}

\end{algorithmic}
\caption{Indicator collection with  lazy validation \label{algo:collect}}
\end{algorithm}


The scaling and contextualization functions used in lines 
8 and 9 of Algorithm \ref{algo:collect} are as follows.

\paragraph{Scaling}

We note that the choice of scaling technique
matters both for classification 
\cite{DBLP:journals/asc/AmorimCC23} and clustering \cite{wongoutong2024impact},
with Percentile transformation among those performing best. 
In what follows, we use percentile transformation, defined by:
for $x$ a value and $V$ a set of values, 
$percentileScaling(x,V)= \frac{|\{x' \in V | x'\leq x \}| \times 100}{|V|}$.

\paragraph{Contextualization}

Once normalized, we use a simple  attenuation coefficient for  indicator
$\langle path, prop, type, op \rangle$ 
based on the length of \textit{path}:
$\frac{1}{1+k}$, where 
$p=(n,\ldots,e_k,n_k)$.

\patrick{Note that Algorithm \ref{algo:collect} lazily evaluates the validation-properties of 
Section \ref{sec:validating} once the set of
candidate indicators is obtained.
However, some validation-properties can be evaluated as early as possible.
For instance, acceptable density 
and discarded properties can be checked when
indicators are devised from properties
(line 2 of Algorithm \ref{algo:collect}).
We therefore implemented a second version of 
Algorithm \ref{algo:collect}, called \textit{Indicator collection with eager validation}, where  acceptable density 
and discarded properties are
pushed down to the indicator collection step.
We postulate that this version, that discards properties early, may speed up the indicator collection and validation.
}



\subsection{Computing comparison insights}

Solving the problem presented in Section \ref{sec:problem} requires to coincidentally produce a partition of both indicators and nodes.
Computing an exact solution to this problem can be achieved using quadratic integer programming or constraint programming, 
but is likely to be intractable on data sets of more than a few hundred of nodes  \cite{Agoston2024, Gilpin_Nijssen_Davidson_2013}.
In this  section, we focus on  practical heuristics that rely on splitting the problem into two optimization stages: first, the partitioning of the indicators, and then the partitioning of nodes.
This has the combined advantages of reducing the combinatorial effort and enabling the use of existing algorithms with few modifications.  
We first present the heuristics for partitioning the set of indicators.

\subsubsection{Exponential heuristic}\label{sec:exponentialHeuristic}

The first heuristic enumerates all partitions of the set of validated indicators  and then solves the nodes partitioning problem, for each partition. This heuristic complexity is $O(2^{|I_{ns}|})$ and is only practical in data sets containing a few indicators, although it is expected to provide the best overall solutions.

\subsubsection{Laplacian based heuristic}

Algorithm \ref{algo:poly-heuristics} gives a simple  heuristic based on analyzing the indicator values, to quickly obtain an approximate solution to the problem.
The idea consists of treating each indicator separately and deciding for them if they will be used for \vero{grouping} 
($P_D$), comparison ($P_S$), or unused \vero{($P_U$)}, based on a classical score representing the goodness for being in one subset or the other.

\begin{algorithm}
\begin{algorithmic}[1]
\Require a set $I$ of indicators for node type $n_S$ with $values(n_S)$ 
\Ensure  \( P= \{P_{S},P_{D},P_{U}\}\) a 3-partition of \(  \mathcal{I}    \)

\State $P_{S}=\emptyset,P_{D}=\emptyset,P_{U}=\emptyset$

\For{$i\in I$} 
\State compute  $LS(i)$ 
\EndFor

\State $Isort=$ sort $I$ by $LS(i)$ 

\State $cut=elbow(Isort)$

\For{$i\in I$} 

\If{$LS(i)<cut$} 

\State $P_D = P_D \cup \{i\}$ \Comment{used for \vero{grouping}} 
\EndIf
\EndFor

\State $I=I\setminus P_D$ 

\For{$i\in I$} 
\State compute  $CV(i)$ 
\EndFor

\State $Isort=$ sort $I$ by $CV(i)$ 

\State $cut=elbow(Isort)$
\For{$i\in I$} 

\If{$CV(i)<cut$} 

\State  $P_U= P_U \cup \{i\}$ \Comment{unused}

\Else

\State $P_S= P_S \cup \{i\}$ \Comment{used for comparison}

\EndIf

\EndFor


\State \Return{$\{P_{S},P_{D},P_{U}\}$}

\end{algorithmic}
\caption{Laplacian heuristic \label{algo:poly-heuristics}}
\end{algorithm}






















Specifically, we use the Laplacian score (LS)  \cite{DBLP:conf/nips/HeCN05} to evaluate the goodness of the indicator to be part of $P_D$, i.e., to be used for clustering.
In clustering, LS is an unsupervised feature selection method that quantifies a feature's ability to preserve the local structure of data by maintaining the proximity of similar data points. It works by constructing a nearest neighbor graph using the feature' values, and computing the graph Laplacian matrix $L = D - A$, where $D$ is the diagonal degree matrix of the graph and $A$ is the similarity 
matrix capturing the connectivity structure of the graph. 

In our case, for each indicator $i$, given a  node type $n_S$, the Laplacian Score is computed as:
$   LS(i) = \frac{I^T L I}{I^T D I}$
where $I=value(n_S,i)$ is the vector of indicator values for $n_S$.

We also use the coefficient of variation $CV$  as a diversity score for choosing comparison indicators.
Given an indicator $i$ for a node type $n_S$, 
it is computed as $CV(i)=\frac{\operatorname{stdev}(value(n_S,i))}{\operatorname{mean}(value(n_S,i))}$.

Algorithm \ref{algo:poly-heuristics} 
first computes LS for each indicator (line 3), sorts the indicators by LS values (line 5) and splits the set of indicators using elbow detection (line 6), keeping those with lowest scores for $P_D$ (lines 7-11).
The same principle is applied for the remaining indicators (line 12), where CV values (line 14) are used to split (line 19) them among $P_U$ (line 20) and $P_S$ (line 22).


\subsubsection{Simple local search}

The  Laplacian heuristic presented above is used as a starting point for a local search, using switching indicators between $P_{S}$, $P_{D}$ and $P_U$ in a greedy manner (see Algorithm \ref{algo:localsearch}).

\begin{algorithm}
\begin{algorithmic}[1]
\Require a set $I_{n_S}$ of indicators with $value(n_S)>$, 
a threshold $\tau$ on the objective function
\Ensure  $P= \{P_{S},P_{D},P_{U}\}$ a 3-partition of $I$

\State $P=$ \vero{Laplacian heuristic} 
($I_{n_S},value(n_S)$) 

\State compute $obj$ for $P$ \Comment{objective function of Eq. \ref{eq:obj}}

\State $obj_{last}=0$

\While{$obj-obj_{last} \geq \tau$} \Comment{stops if gain below $\tau$}

\State $obj_{last}=obj$


\For{all permutations $perm(P)=P'$} 

\State compute $obj_{new}$ for $P'$ \Comment{objective function of Eq. \ref{eq:obj}}

\If{$obj_{new}>obj_{last}$} 

\State $obj=obj_{new}$

\State $P_{best}=P'$

\EndIf

\EndFor

\State $P=P_{best}$ \Comment{new $P$ is the best 3-partition}

\EndWhile

\State \Return{$P$}

\end{algorithmic}
\caption{Greedy local search \label{algo:localsearch}}
\end{algorithm}


More precisely, let $perm(P)=P'$ be a permutation operation that changes $P= \{P_{S},P_{D},P_{U}\}$ into
a new 3-partition $P'= \{P'_{S},P'_{D},P'_{U}\}$ 
such that $|P_{S}\setminus P'_{S}| + |P_{D}\setminus P'_{D}| + |P_{U}\setminus P'_{U}| =1$.
Algorithm \ref{algo:localsearch} computes a first solution $P$ to the problem with Algorithm \ref{algo:poly-heuristics} (line 1).
The quality of this 3-partition is computed as the score of the objective function of Equation \ref{eq:obj}, i.e., the score given by the function to maximize. 
It then iterates while the gain in the objective function remains above a user-specified threshold (line \vero{4).} 
At each iteration, all permutations, swapping two indicators from the subsets of $P$ from the current solution, are computed (line 7), and the one with the maximum gain (lines 9-13) is used for the next iteration.

\subsubsection{Clustering heuristic for node partitioning}

We use  a modified clustering algorithm to partition the set of nodes once the partitioning of indicators is decided. Two approaches can be considered. The first approach would consist of treating the first and the second term of the objective (see Equation \ref{eq:obj}) separately, thus obtaining two partitions of nodes in $k$ clusters. Those two partitions can subsequently be combined. This is known as consensus clustering \cite{10.1109/ICDM.2007.73} and mainly used when no access to the underlying data is available.
In our case, since data are known, we chose instead  to compute a single partition. We use a modified version of a fuzzy c-medoids algorithm \cite{fcmd} to consider both terms of our objective in the distance computation (instead of only in the second term as is classically done). 
%
Fuzzy clustering offers improved handling of outliers and enables a more gradual exploration of the cluster center solution space compared to crisp algorithms such as k-means. In particular, fuzzy c-means assigns each instance a degree of membership across clusters, encoded in a membership matrix. This matrix is then used to compute weighted averages for the cluster centers, allowing the optimization process to better capture the underlying structure of the data. As a result, the clustering component of the loss function is typically more finely optimized than in traditional crisp approaches, albeit at the cost of increased spatial complexity due to the membership matrix.

\subsubsection{Random + local search heuristic}

Finally, we  also modified  Algorithm \ref{algo:localsearch} by starting from a random distribution of the indicators of $I$ in $P$, performing the greedy local search, clustering nodes, and repeating this process 5 times (i.e., from 5 different random distributions). We retain the  best solution regarding the score obtained by Equation \ref{eq:obj} among the 5 obtained.

Note that the algorithmic solutions we propose in this Section compute clusters of nodes, given by a comparison scheme and a grouping scheme, and not directly the comparison insights per se. From the obtained clusters, comparison insights can easily be presented cluster-wise, by giving pairs of nodes for which the significance of the indicators in the comparison scheme is the highest.


%% file: tests.tex
\section{Tests}\label{sec:tests}

\subsection{Experimental setup}

We use python 3.10 and Neo4j 5.24 with  APOC 5.24.2.
In our implementation, most of the computation for collecting candidate indicators is delegated to Neo4j through Cypher queries.
We only consider numerical properties for devising indicators, and sum as the aggregation function.

\subsubsection{Data sets}

\begin{table}[]
    \centering
    \begin{tabular}{lrrrrrr}
    &&&&& \multicolumn{2}{c}{Avg \vero{nb.} prop
    } \\
    \cline{6-7}  
    Database        &  $|N|$      & $|E|$       &$|N_S|$&$|E_S|$&  N    & E\\
         \hline 
    Airports    & 52,944    & 136,948   & 3     & 5     & 4.39  & 2.13\\
    ICIJ leaks  & 2,016,523 & 3,339,267 & 5     & 69    & 1.00     & 0.00\\
    Movies      & 28,863    & 166,261   & 6     & 4     & 1.60   & 1.20\\
    \end{tabular}
    \caption{Graph data sets used in the tests}
    \label{tab:datasets}
\end{table}

The data sets used in our tests are summarized in Table \ref{tab:datasets}.
We include real graph data set from various sources.
The Movie data set is the recommendation graph 
from the Neo4j example repository\footnote{\url{https://github.com/neo4j-graph-examples}}.
The ICIJ leaks data set is from the International Consortium of Investigative Journalists\footnote{\url{https://offshoreleaks.icij.org/}}.
Finally, to test over a graph where nodes have many numerical properties, we created a Neo4j database (Airports) using various sources collected on Kaggle.com and Eurostat\footnote{\url{https://ec.europa.eu/eurostat/web/main/home}}, and data scrapped from Google Flights using the fast-flights library\footnote{\url{https://aweirddev.github.io/flights/}}. We developed an ETL process to unify and standardize the data, ensuring consistency and integrity.
For these databases, all numerical properties in nodes or edges were indexed with range indexes.
We also include a classical UCI data sets (Iris), chosen for
its very small size (4 columns, simulating indicators, and 150 rows, simulating the nodes to compare).

\subsubsection{Protocol}

We assess our algorithms' efficiency (run time in seconds) over the data sets, as well as the quality of the different heuristics, by observing the score of the objective function (Equation \ref{eq:obj}). We compare the efficiency and score of the heuristic to a \textit{random} baseline, where the indicators are randomly split between $P_S, P_D$ and $P_U$. The number of clusters vary in $[2,4]$.

Testing indicator collection was done on a Macbook Pro laptop equipped with an Apple Silicon  M2 Pro and 32GB of RAM, running MacOS 15.4. 
The more computationally demanding comparison insight extraction experiments were conducted on a Fedora Linux workstation, with two 2.3 GHz Intel Xeon 5118 12-core CPUs and 377GB 2666 MHz of DDR4 main memory.

\subsection{Runtime for collecting indicators}

\begin{figure}
    \centering
    \includegraphics[width=8cm]{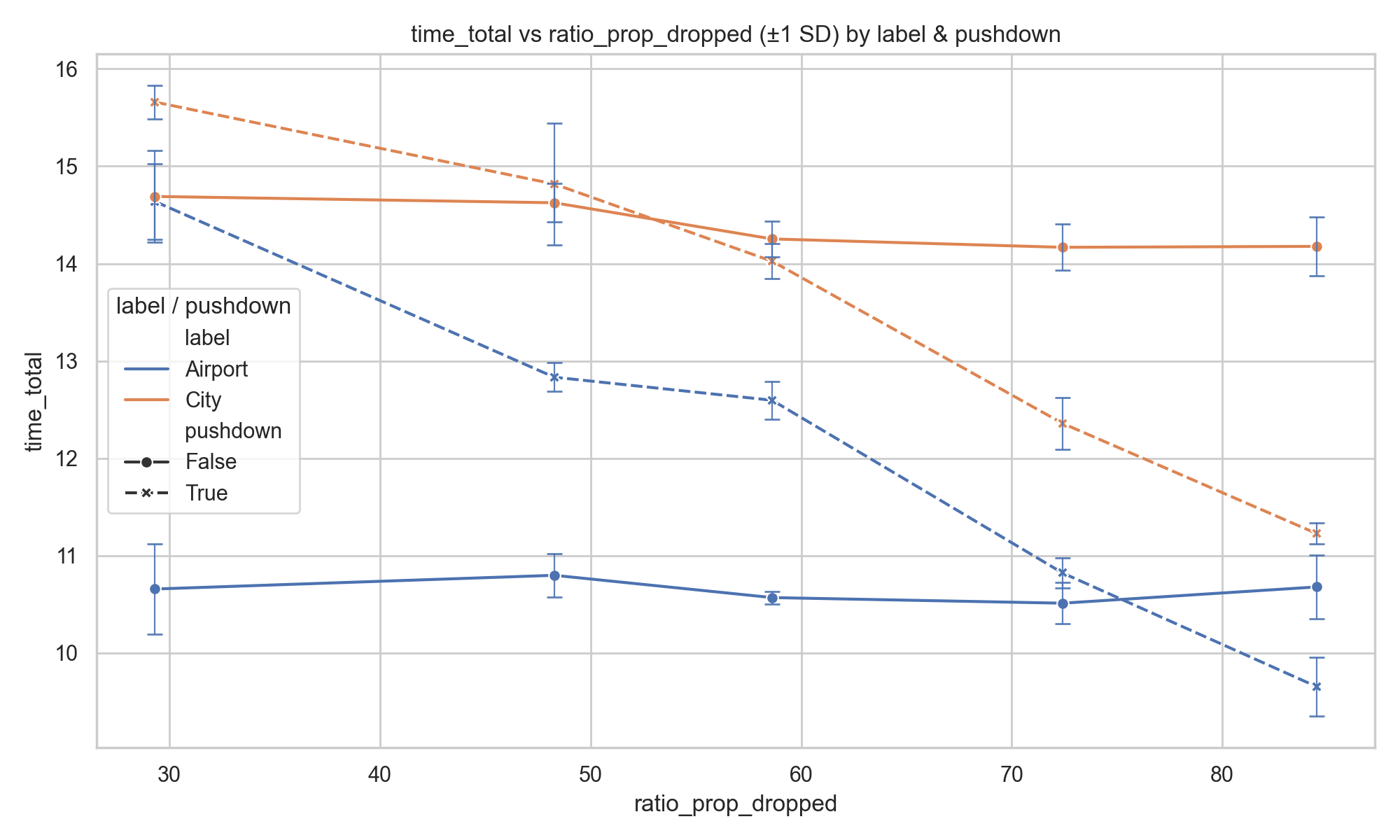}
    \caption{Lazy vs eager validation (total time (s))}
    \label{fig:lazy-vs-eager-validation-total}
\end{figure}

\begin{figure}
    \centering
    \includegraphics[width=8cm]{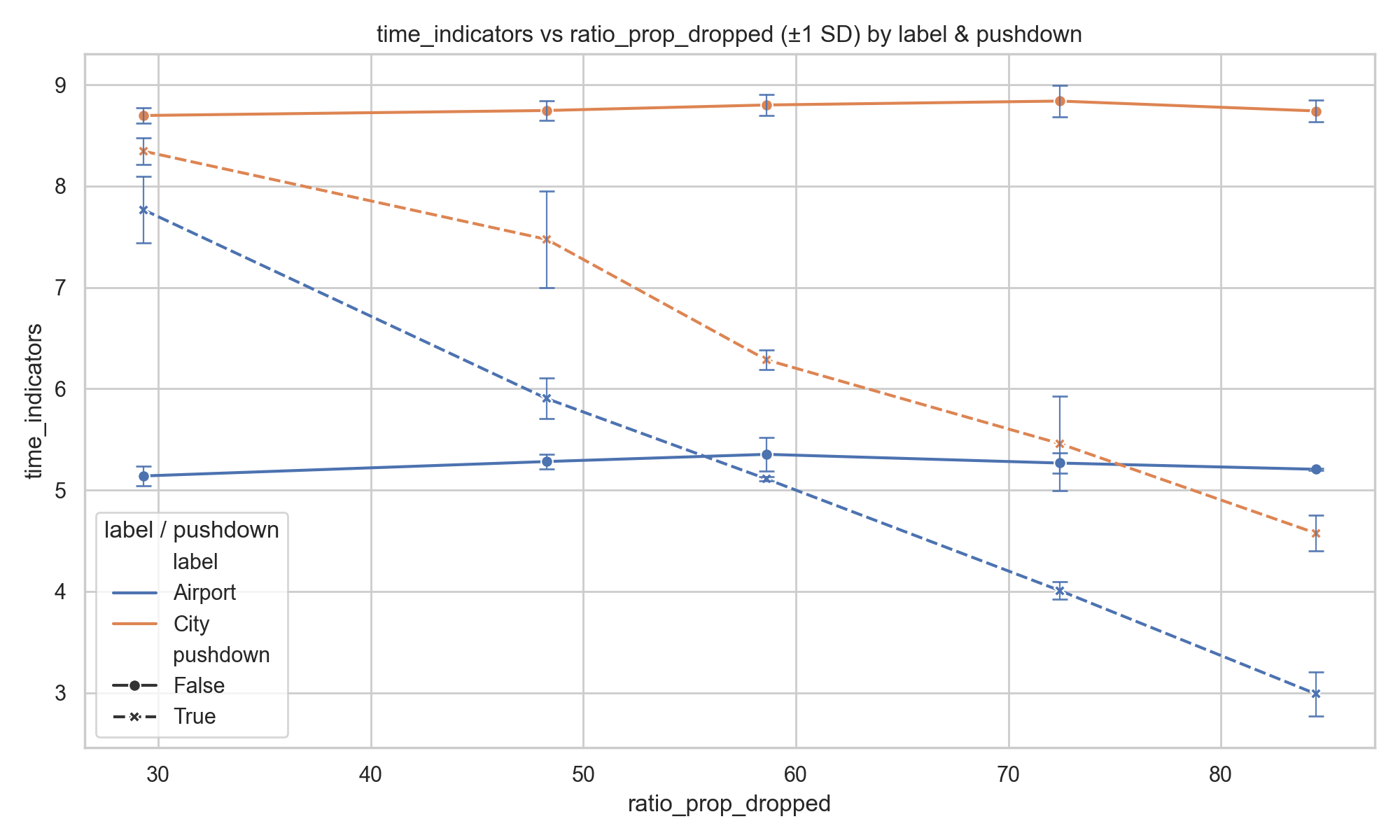}
    \caption{Lazy vs eager validation (candidate indicator collection time (s))}
    \label{fig:lazy-vs-eager-validation-collect}
\end{figure}

We first start by analyzing the efficiency of the two versions, with lazy (pushdown=False) or eager (pushdown=True) validation, of Algorithm \ref{algo:collect}. 
The test consists of collecting indicators varying the number of properties to discard (by changing the acceptable density threshold), and observing the time taken by the two algorithms, averaged over 3 runs. We used  the Airports database, with two labels ($n_S=$ Airport or City) chosen to illustrate the impact of size (44,997 nodes for City against 7,698 for Airport).
It can be seen on Figure \ref{fig:lazy-vs-eager-validation-total} that the more properties are discarded,
the more eager validation is efficient. 
The impact is more important when the number of nodes to explore for collecting indicators is higher.
As Figure \ref{fig:lazy-vs-eager-validation-collect} indicates, the speedup is important for the collection step (line 2 in Algorithm \ref{algo:collect}), since this step relies on Cypher queries, for which push-down results in queries being more selective, while validation (lines 4-9) is done in main memory.

{\footnotesize
\begin{table}[!ht]
 \begin{tabular}{llllllll}
\toprule
&&& \multicolumn{4}{c}{time (s)} \\
\cline{3-7}
Database &  $n_S$ & $|IE(n_S))|$ &  
Card. & Cand. & Valid. & total \\
\midrule
Airports & Airport & 7,698 &  
0.8 & 5.1 & 2.3 & 10.4\\
Airports & City &   44,997 & 
0.8 & 8.8 & 2.8 & 14.6\\
Airports &  Country &  2,49& 
0.8 & 1.2 & 2.2 & 6.5 \\
ICIJ leaks & Entity & 814,344 & 
94.7 & 210.4 & 58.4 & 364.3 \\
ICIJ leaks  & Interm. & 26,768  & 
94.7 & 13.3 & 54.1 & 162.8 \\
ICIJ leaks &  Officer & 771,315 & 
94.7 & 236.9 & 56.9 & 389.2\\
Movies & Actor &  15,443 & 
0.8 & 5.4 & 3.6 & 10.5  \\
Movies &  Director & 4,091 & 
0.8 & 1.6 & 3.4 & 6.6\\
Movies & Movie &  9,125 & 
0.8 & 2.7 & 3.5 & 7.8\\
\bottomrule
\end{tabular}
    \caption{Runtime breakdown for collecting indicators (lazy validation, acceptable density = 80\%)}
    \label{tab:timeP1}
\end{table}
}



Table \ref{tab:timeP1} details the time (averaged over 3 runs) needed to obtain $values(n_S)$ from various $n_S$ of the 3 graph data sets,
that is computing the cardinalities of relationships in the graph and then collecting and validating indicators (Algorithm \ref{algo:collect}, with lazy validation). 
$IE(n_S)=\Inst(\type(n_S))$.
\textit{Card.} is the time for computing relationship cardinalities, 
\textit{Ind.}, the time for computing candidate indicators and 
\textit{Val.}, the time for validating the set of indicators. 

Validation of the indicators is made with: acceptable variance ratios of $10^{-4}\%$ for $\alpha$, $100\%$ for $\beta$, acceptable density of $80\%$, Pearson correlation coefficient $>0.98$ for non redundancy, and discarding identifiers.
These thresholds were empirically chosen to remove meaningless indicators.
Nulls  were dropped from $values(n_S)$ to ensure that significance and distance can be computed.

It can be seen that, expectedly, the time to compute cardinalities depends on $|E|$, the number of edges of the graph (see Table \ref{tab:datasets}) 
while the time to compute the candidate indicators is mostly impacted by$|IE(n_s)|$. More surprisingly, the validation time depends not only on $|IE(n_s)|$ but also on $|E|$ (Spearman correlation coefficient of 0.95), which is explained by the fact that the larger $|E|$, the more likely it is to have larger $|I_{n_S}|$ due to context being used for computing new indicators for $IE(n_s)$.

\subsection{Runtime for computing comparison insights}

{\footnotesize
\begin{table}[!ht]
    \centering
 \begin{tabular}{llrr}
\toprule
   & & Nb. validated & Nb. retained  \\
Data set & $n_S$ & indicators & nodes  \\
\midrule
Iris & - & 4 & 150    \\
Movies & Director & 10 & 730   \\
Movies & Movie & 10 & 4143    \\
Airports & Airport & 28 & 3837 \\
\bottomrule
\end{tabular}
    \caption{Sets of indicators for insight extraction}
    \label{tab:datasets-for-clustering}
\end{table}
}


This series of tests reports the time taken to extract comparison insights once the indicators are validated.
For these tests, the data used are described in Table \ref{tab:datasets-for-clustering} and the heuristics compared are: random (\textit{rd}), Laplacian (\textit{lp}), simple local search (\textit{sls}), random + local search (\textit{ls}).
Each test is run 10 times and the averages and standard deviations are reported.



\begin{figure}
    \centering
    \includegraphics[width=7cm]{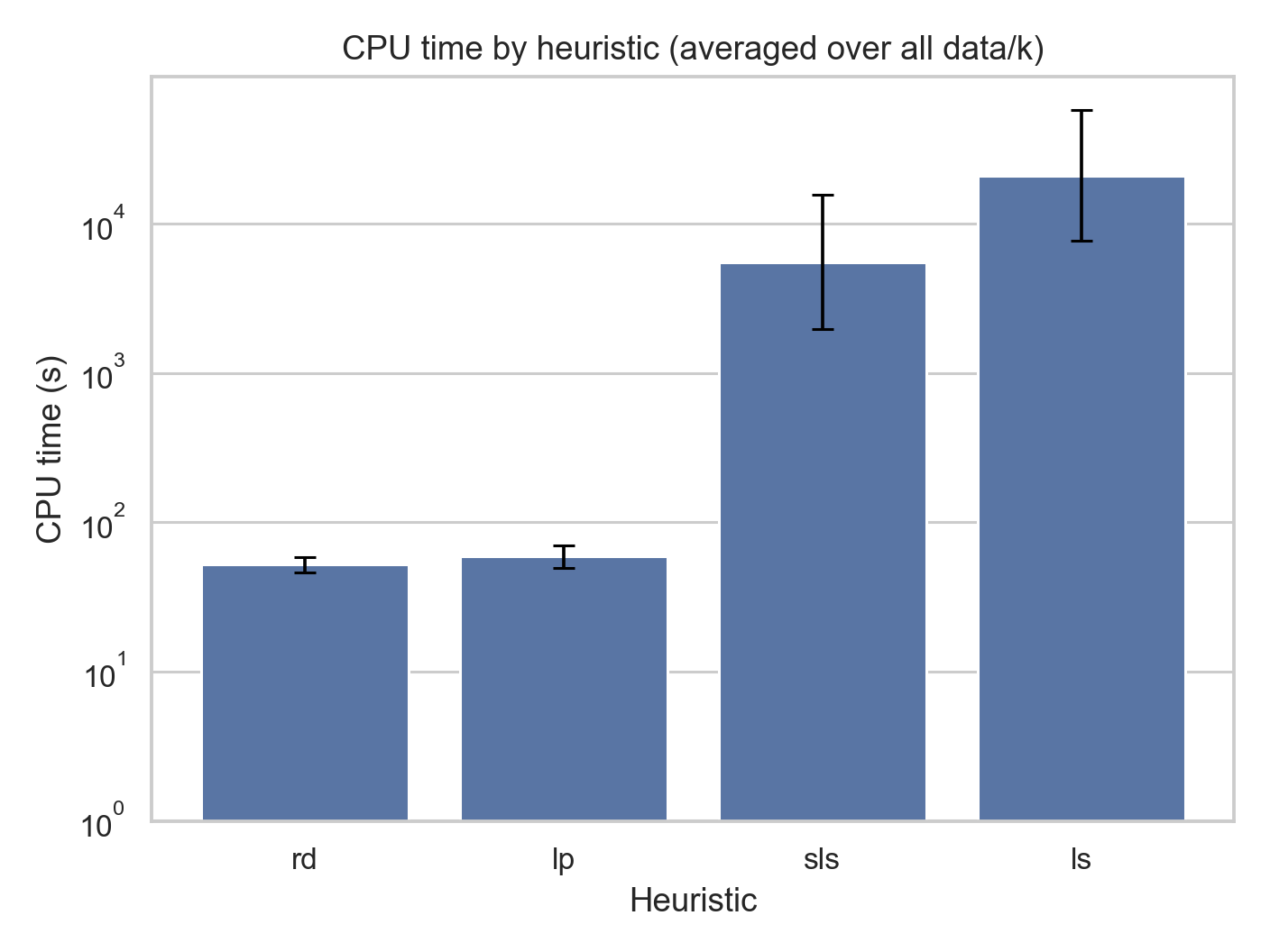}
    \caption{CPU time of the 4 heuristics}
    \label{fig:cpu}
\end{figure}

Figure \ref{fig:cpu}
reports the raw CPU time, considering the program as single threaded,
for all heuristics. 
Expectedly, random and Laplacian heuristics are the fastest, while both local search heuristics are slower, with \textit{ls} the slowest due to the repetition of the search
for a better exploration of the search space.

\begin{figure}
    \centering
    \includegraphics[width=7cm]{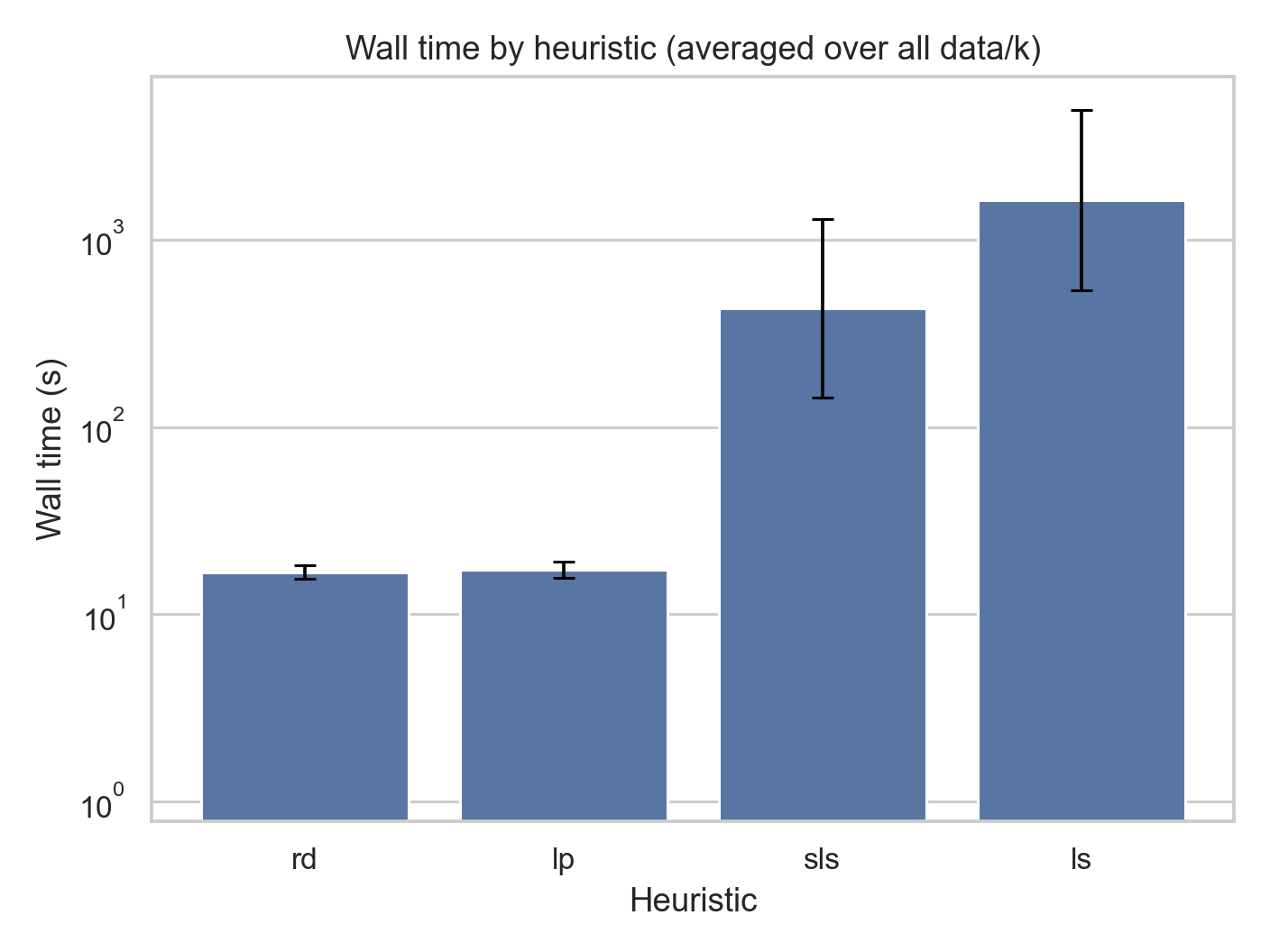}
    \caption{Wall time of the 4 heuristics}
    \label{fig:thewall}
\end{figure}

We also report in Figure \ref{fig:thewall} the time taken when multi-threading is used.
It can be seen that all the heuristics can 
take advantage of the mutli-core processor, since the Wall time is about one order of magnitude lower than
 the CPU time.

\begin{figure}
    \centering
    \includegraphics[width=7cm]{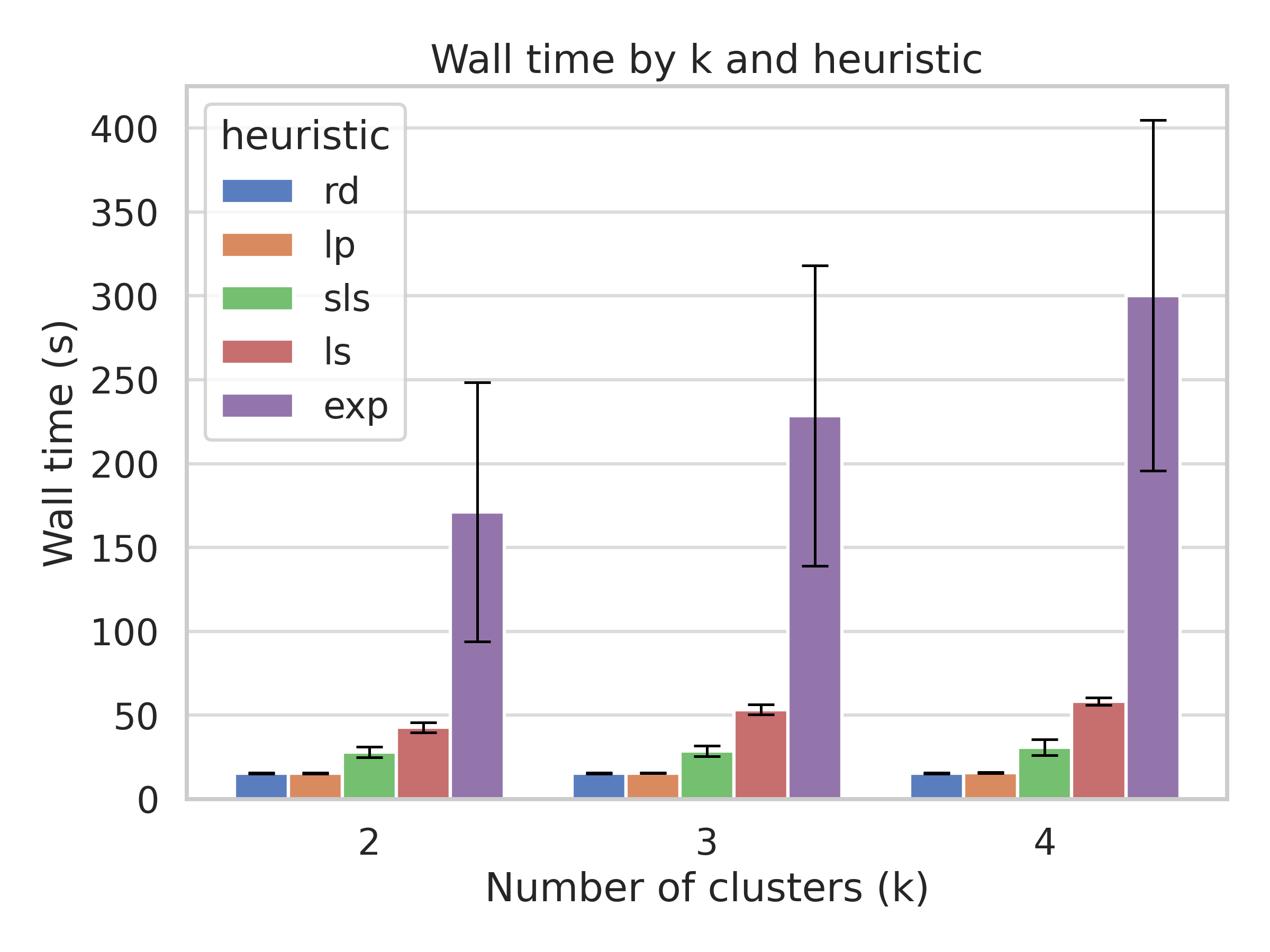}
    \caption{Wall time by number of clusters (Iris)}
    \label{fig:wall-iris}
\end{figure}

Figure \ref{fig:wall-iris} shows the Wall time of all heuristics, including the exponential one (\textit{exp}), on the small Iris data set, analyzing the influence of the desired number $k$ of clusters.
It can be seen that the naive heuristics (random, Laplacian) are   insensitive to $k$, while \textit{ls, sls} and \textit{exp} are impacted by larger $k$. More importantly, on this small data set, exponential is faster than \textit{ls}, which is expected since the search space is small and \textit{ls} suffers from a fix number of repetitions.
For comparison, \textit{exp}  timed out on Movie after 10 hours for $k=2$.

\subsection{Quality of the solution found}




\begin{figure}
    \centering
    \includegraphics[width=9cm]{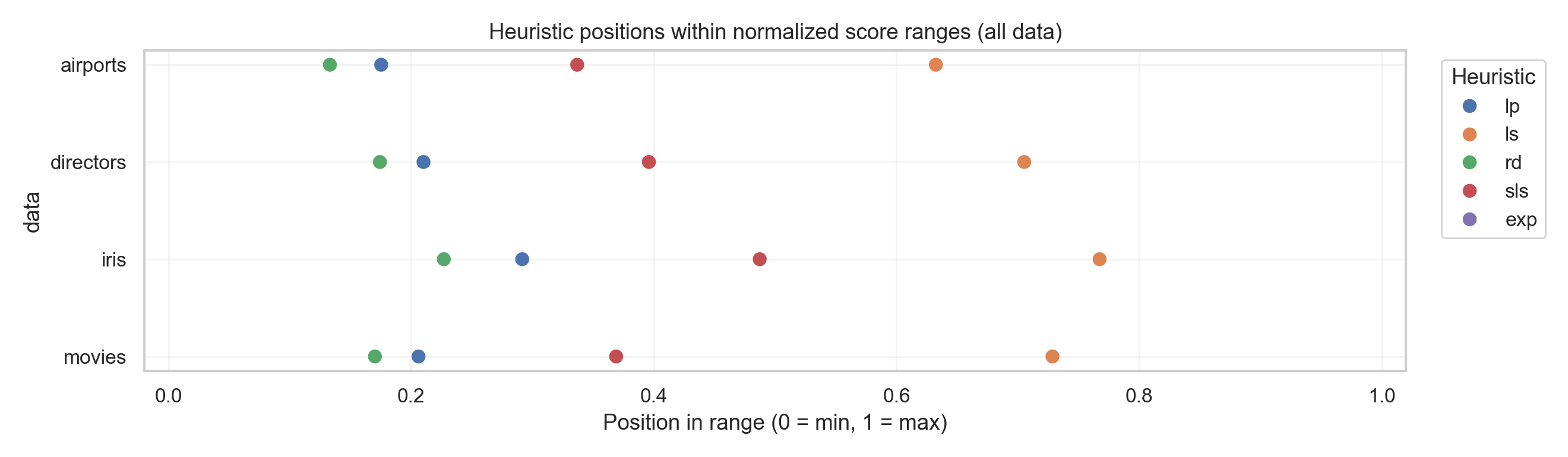}
    \caption{Objective score range}
    \label{fig:scoreRange}
\end{figure}

In this last series of tests, we analyze the quality of the insights found. This quality is expressed in terms of the score obtained for the objective function (Equation \ref{eq:obj}).
Figure \ref{fig:scoreRange}  reports
the normalized range of scores by data set, and plots the score obtained by each heuristic. 
Normalization is done by considering all the scores obtained on a data set by each heuristic across all runs, and using min-max to normalize.
It can be seen that $ls$ achieves the best scores, which is explained by the fact that it has 5 chances to reach a good solution using local search. 
Unsurprisingly, Random and Laplacian heuristic score the worst, which motivates the use of local search.

\begin{figure}
    \centering
    \includegraphics[width=7cm]{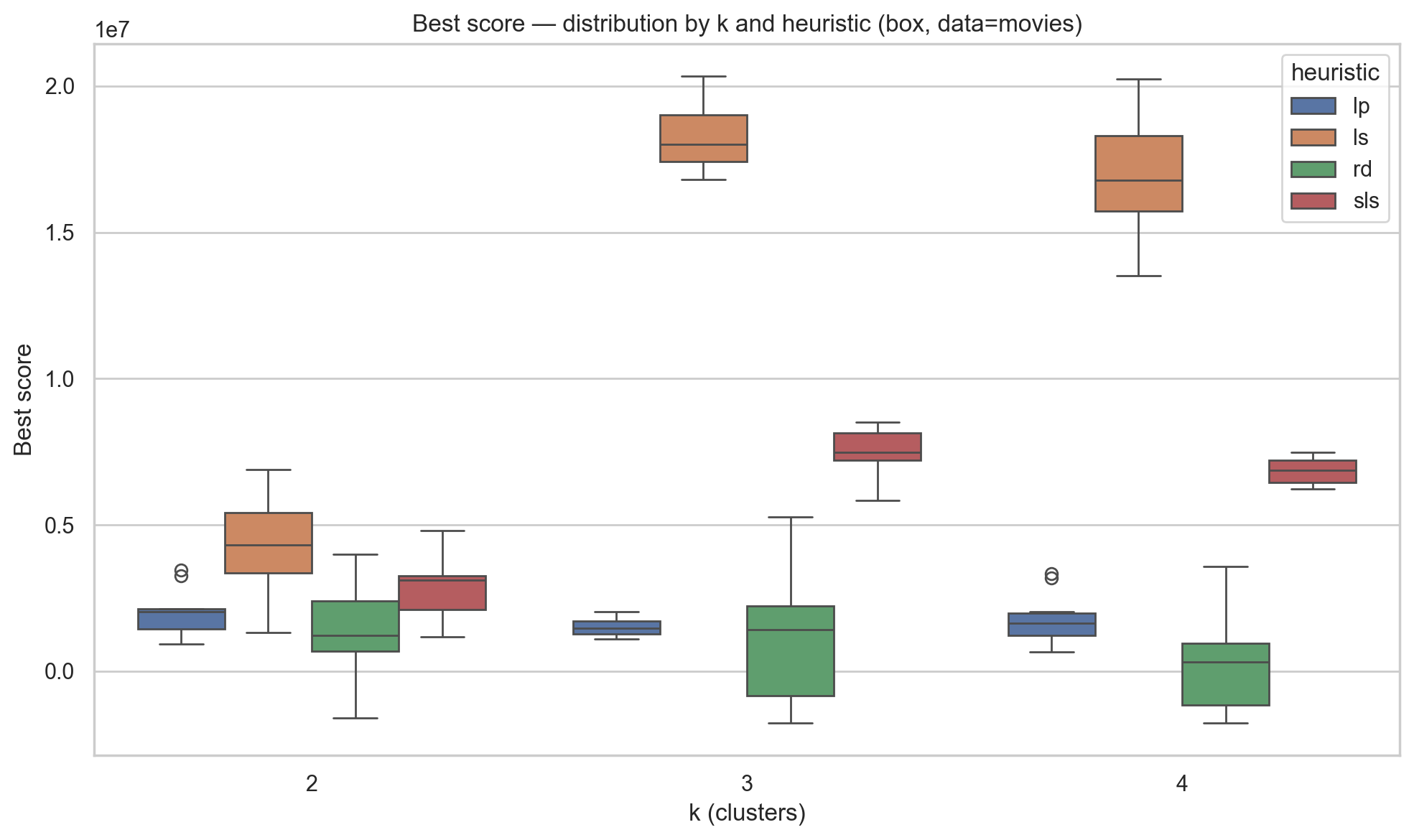}
    \caption{Objective score by heuristic and number of clusters (Movie)}
    \label{fig:scoreByK-movies}
\end{figure}

\begin{figure}
    \centering
    \includegraphics[width=7cm]{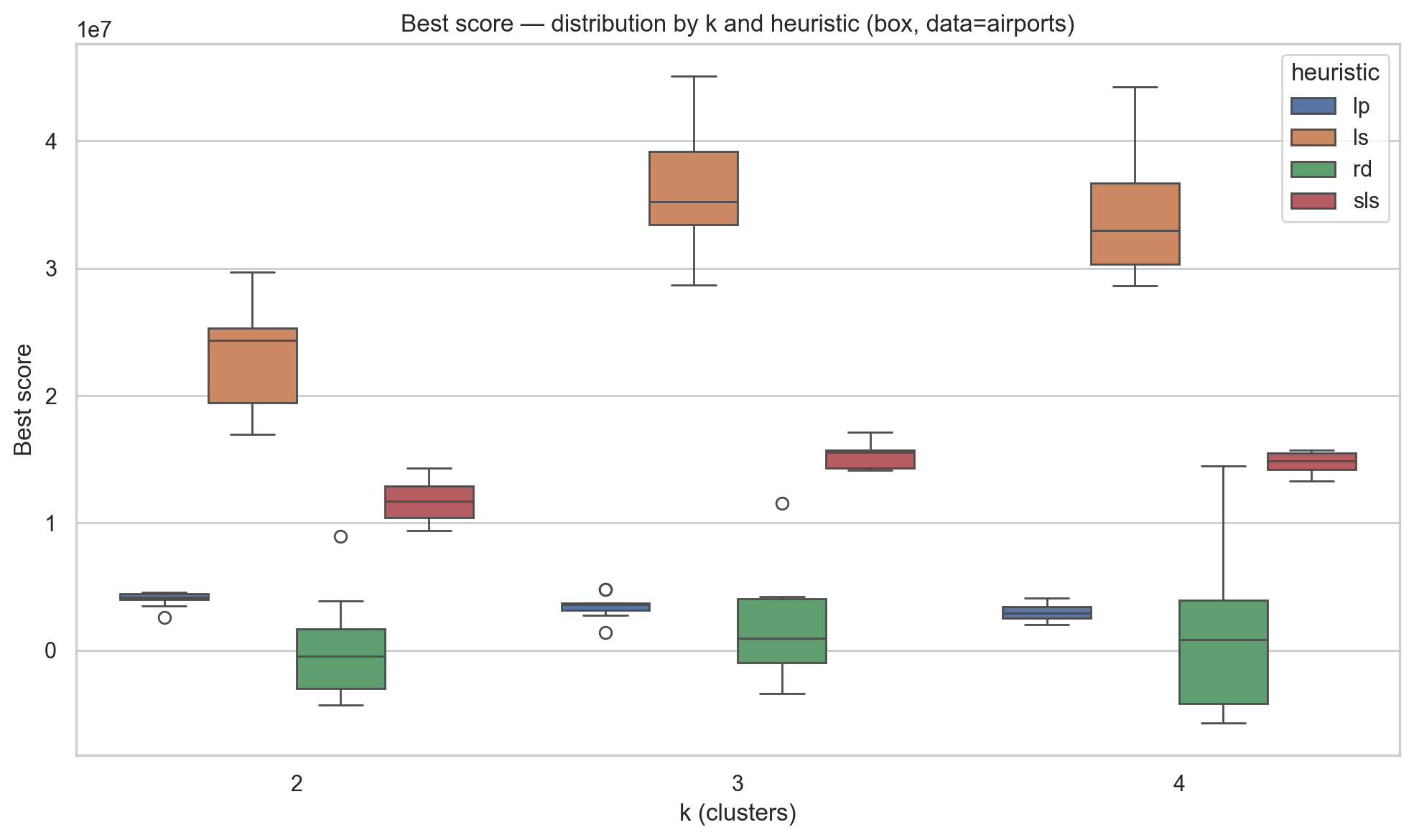}
    \caption{Objective score by heuristic and number of clusters (Airport)}
    \label{fig:scoreByK-airports}
\end{figure}

\begin{figure}
    \centering
    \includegraphics[width=7cm]{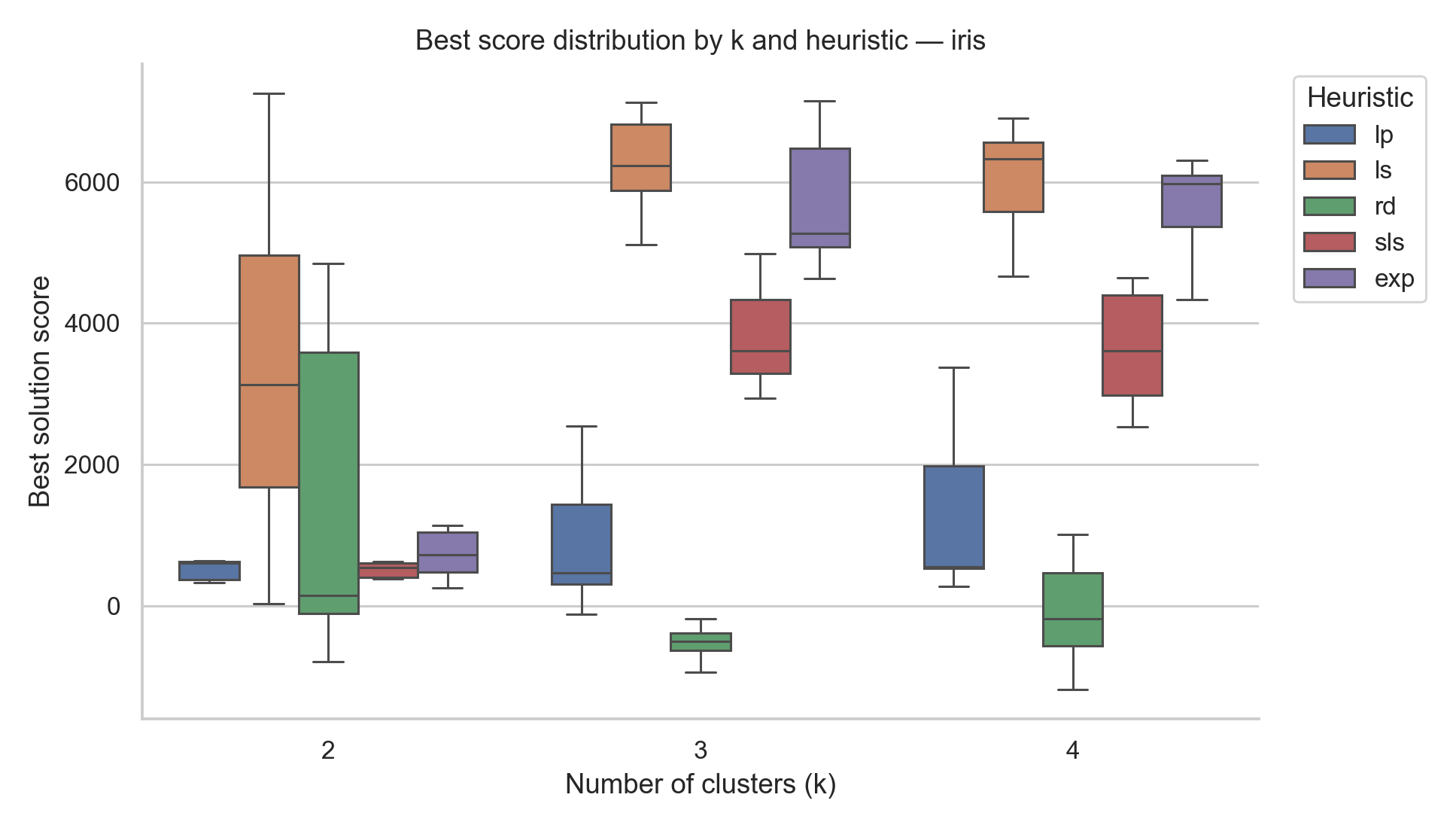}
    \caption{Objective score by heuristic and number of clusters (Iris)}
    \label{fig:scoreByK-iris}
\end{figure}

Figures \ref{fig:scoreByK-movies} and \ref{fig:scoreByK-airports} display the distributions of scores for 2 node types (Movie for the Movies database and Airport for the Airports database), by heuristic and number of cluster.
The tendency is that higher $k$'s favor a better quality, but a plateau seems to be reached for $k=4$.
Figure \ref{fig:scoreByK-iris} shows the score obtained by each heuristic, including the exponential one (\textit{exp}), on the small Iris data set. 
It can be seen that, while \textit{exp} does not achieve the best scores, which is expected on this small data set, it is less sensitive to this plateau phenomenon, indicating the interest of a better exploration of the search space when increasing $k$.




%% file: related.tex
\section{Related work}\label{sec:related}

Our paper formalizes the problem of automatically extracting meaningful comparisons between nodes in property graphs based on their context and introduces heuristics to solve it efficiently. This situates our work within Exploratory Data Analysis (EDA), the often tedious task of interactively analyzing datasets to gain insights —an area that has attracted attention both recently \cite{DBLP:conf/sigmod/IdreosPC15,milo_2020_sigmod_tuto} and since the early days of discovery-driven exploration 
of multidimensional data ~\cite{DBLP:conf/edbt/SarawagiAM98}. Approaches for automatically generating insights to support EDA generally fall into two categories: \textit{generate-and-select} methods \cite{DBLP:conf/sigmod/DingHXZZ19,DBLP:conf/sigmod/TangHYDZ17}, which enumerate and rank potential insights, and \textit{guided EDA} methods \cite{DBLP:conf/sigmod/ElMS20,eda4sum}, which iteratively explore the search space in a human-like manner. Our approach belongs to the former category, focusing on generating and prioritizing meaningful comparisons between nodes based on their contextual relevance.
 
Several studies highlighted the importance of comparisons
when analyzing data \cite{DBLP:conf/chira/BlountKZS20,DBLP:conf/chi/ZgraggenZZK18}.
Approaches exist to extract comparison insights \cite{DBLP:conf/edbt/ChansonLMRT22} 
or express comparison queries \cite{DBLP:journals/pvldb/SiddiquiCN21},
though, to our knowledge, only in relational databases. Our work, in contrast, targets analytical tasks on property graphs.

Graph analytics generally refers to analyzing and enriching graph data beyond basic OLTP queries, leveraging structural patterns to enable advanced analyses \cite{DBLP:journals/sigmod/BonifatiOTVYZ24,cheramangalath2020introduction}. While classical algorithms such as PageRank and ShortestPaths remain central in benchmarks like LDBC Graphalytics\cite{iosup2016ldbc}, recent research focuses on extending application-driven functionalities that allow users to perform sophisticated analyses.  For example, recent work has introduced streaming operators in Cypher \cite{rost:hal-04798351} and causal inference constructs in GQL \cite{10.14778/3749646.3749671}. These studies, however, differ from our work on their purpose. 

Other works, such as \cite{dumbrava2019approximate}, explore property graph analytics to estimate counting regular path queries. The aim is to aggregate nodes into supernodes to produce statistical properties like the number of inner vertices, rather than examining the properties of individual nodes to generate indicators. \cite{10.1007/978-3-319-68288-4_31} investigates a domain-independent approach to entity comparison over RDF graphs, leveraging queries to compute similarities and differences without relying on explicit paths. While this study generates a single SPARQL query to explain entity similarity, our work derives a set of indicators, providing a comparison of property graphs based on properties and the topology. 

Our approach adopts a multi-stage, filter-based process refined through heuristic search, where features are evaluated based on their intrinsic properties. It comprises two stages: the first dedicated to devising indicators and the second to computing insights. The first stage aligns with feature engineering principles, as the comparison indicators act as features capturing structural and semantic characteristics of the graph. Feature engineering techniques aim to reduce the search space and enhance model performance, mitigating the curse of dimensionality when generating comparative insights from high-dimensional graphs \cite{DBLP:journals/apin/DhalA22,BARBIERI2024123667,9885186}. While the focus of our work on node comparison insights represents a novel contribution, our methodology builds upon well-established practices that employ filter-based selection for devising indicators, together with scaling and attenuation techniques.

The second stage employs well-established heuristic refinement techniques by integrating Laplacian heuristics with random and local search strategies~\cite{9885186,DBLP:conf/nips/HeCN05,DBLP:journals/apin/DhalA22}. This approach harnesses the structural properties of the Laplacian to guide the search process while combining them with the exploratory and fine-tuning capabilities of random and local search techniques \cite{DBLP:books/daglib/0004035}. The result is a solution that can compute insights, even for large graphs, without fully exploring the search space. Related studies that also apply heuristic-based methods to optimize the search space, but have objectives distinct from ours, include the automatic generation of comparison tables from knowledge bases~\cite{DBLP:conf/esws/GiacomettiMS21}, automatic facet generation~\cite{feddoul2019automatic}, and the automatic derivation of ranking indicators for knowledge graphs~\cite{abdallah2024ranking}.

A  key original aspect of our work lies in the definition of indicators for property graph based on the context of nodes, alongside the formalization of our problem and the proposal of several approaches to its solution.

%% file: conclusion.tex
\section{Conclusion}\label{sec:conclusion}

This paper presents an approach to extract node  comparison insights in property graphs.
Comparison insights rely on indicators build from the identification of the context of nodes to be compared. We formalize the problem of choosing which indicators to group nodes and which ones to compare them within groups, and propose various heuristics to obtain approached solutions to this problem.

Our short term perspectives consist of tuning our approach, for instance by injecting user preferences for devising indicators,
handling validation-properties at more granular levels, or
leveraging global properties (like node centrality measures).
On the long run, we plan to investigate other types of comparison insights specific to property graphs (like e.g., path comparisons), and find ways to automatically complement property graphs with additional numeric properties, making them better fit for extracting comparisons.


%% file: ai.tex
\section{AI-Generated Content Acknowledgement}


GPT 5 has been used to implement parts of the algorithms and Cypher queries used for the experiments (see Sections \ref{sec:algo} and 
\ref{sec:tests}). Our code is available in the git repository: \url{https://github.com/AlexChanson/Comparing-Nodes}.